\documentclass[10pt,a4]{article}

\usepackage{graphicx}              
\usepackage{amsmath}               
\usepackage{amsfonts}              
\usepackage{amsthm}                
\usepackage{color}
\usepackage{amsfonts,amsmath,amssymb,graphics,psfrag}
\usepackage{hyperref}

\makeatletter
\@addtoreset{equation}{section}
\makeatother

\DeclareMathOperator{\id}{id}
\newcommand{\nn}{\nonumber}
\newcommand{\RR}{\mathbb{R}}      
\newcommand{\CC}{\mathbb{C}}      

\newcommand{\be}{\begin{eqnarray}}
\newcommand{\ee}{\end{eqnarray}}

\newcommand{\sgn}{\mbox{sgn}}
\newcommand{\dimm}{\mbox{dim}}

\newcommand{\equ}{\stackrel{!}{=}}

\newcommand{\DC}{\mathcal{D}}

\newcommand{\QC}{\mathcal{Q}}
\newcommand{\PC}{\mathcal{P}}

\newcommand{\HC}{\mathcal{H}}
\newcommand{\FC}{\mathcal{F}}
\newcommand{\LC}{\mathcal{L}}
\newcommand{\OC}{\mathcal{O}}
\newcommand{\TC}{\mathcal{T}}
\newcommand{\IC}{\mathcal{I}}
\newcommand{\AK}{\mathfrak{A}}

\newcommand{\braket}[1]{\left\langle #1 \right\rangle }
\newcommand{\braaket}[2]{\left\langle #1\mid #2 \right\rangle }
\newcommand{\braaaket}[3]{\left\langle #1\mid #2 \mid #3 \right\rangle }

\newcommand{\ket}[1]{\left| #1 \right\rangle}

\begin{document}

\title{{\sf Born--Oppenheimer Decomposition for Quantum Fields on 
Quantum Spacetimes}}
\author{
{\sf Kristina Giesel$^{4,5}$}\thanks{{\sf 
kristina.giesel(at)aei(dot)mpg(dot)de, giesel(at)nordita(dot)org}},
{\sf Johannes Tambornino$^{1}$}\thanks{{\sf 
johannes(dot)tambornino(at)aei(dot)mpg(dot)de}},
{\sf Thomas Thiemann$^{1,2,3}$}\thanks{{\sf 
thiemann(at)aei(dot)mpg(dot)de, 
tthiemann(at)perimeterinstitute(dot)ca, 
thiemann(at)theorie3(dot)physik(dot)uni-erlangen(dot)de}}\\
\\
{\sf $^1$ MPI f. Gravitationsphysik, Albert-Einstein-Institut,} \\
           {\sf Am M\"uhlenberg 1, 14476 Potsdam, Germany}\\
\\
{\sf $^2$ Universit\"at Erlangen, Institut f\"ur Theoretische Physik 
III, Lehrstuhl f\"ur Quantengravitation} \\
           {\sf Staudtstr. 7, 91058 Erlangen, Germany}\\           
\\
{\sf $^3$ Perimeter Institute for Theoretical Physics,} \\
{\sf 31 Caroline Street N, Waterloo, ON N2L 2Y5, Canada}\\
\\
{\sf $^4$ Nordic Institute for Theoretical Physics (NORDITA),} \\
{\sf Roslagstullsbacken 23, 106 91, Stockholm, Sweden}\\
\\
{\sf $^5$ Excellence Cluster 'Universe', Technische Universit\"at 
M\"unchen,} \\
{\sf Boltzmannstr. 2, 85748 Garching, Germany}
}
\date{}

\maketitle
\thispagestyle{empty}

\begin{abstract}
{\sf 
Quantum Field Theory on Curved Spacetime (QFT on CS) is a well established 
theoretical framework which intuitively should be a an extremely 
effective description of the quantum nature of matter when propagating 
on a {\it given background spacetime}. If one wants to take care of 
backreaction effects, then a theory of quantum gravity is needed. It is  
now widely believed that such a theory should be formulated in a non --
perturbative and therefore {\it background independent} fashion. Hence,
it is a priori a puzzle how a background dependent QFT on CS should 
emerge 
as a semiclassical limit out of a background independent quantum gravity 
theory.

In this article we point out that the Born -- Oppenheimer decomposition 
(BOD) of the Hilbert space is ideally suited in order to establishing 
such 
a link, provided that the Hilbert space representation of the 
gravitational field 
algebra 
satisfies an important condition. If the condition is satisfied, then 
the 
framework of QFT on 
CS can be, in a certain sense, embedded into a theory of quantum 
gravity. 

The unique representation of the holonomy -- flux algebra 
underlying Loop Quantum Gravity (LQG) violates that condition. While it 
is conceivable that the condition on the representation can be relaxed, 
for convenience in this article we consider a new classical
gravitational field algebra and a Hilbert space representation of its 
restriction to 
an algebraic graph for which the condition is satisfied. An important
question that remains and for which we have only partial answers is 
how to construct eigenstates of the full gravity -- matter 
Hamiltonian whose BOD is confined to a small neighbourhood of a 
physically interesting vacuum spacetime.
}
\end{abstract}

\vspace{2cm}

\newpage


\section{Introduction} \label{sec:introduction}

Quantum Field Theory (QFT) on Curved Spacetime (CS) is nowadays a well
established discipline of mathematical physics (see \cite{Hollands} for 
a recent review and \cite{Verch} for a functorial formulation). It is 
designed 
to describe the 
regime of physics 
in which backreaction effects of matter on geometry and the quantum 
fluctuations of geometry itself can be safely neglected. In order to 
go beyond that, a theory of quantum gravity is needed. It is nowadays 
widely believed that perturbative quantum gravity is non renormalisable 
and therefore a non perturbative formulation has to be found. Of course,
such a theory of quantum gravity must contain QFT on CS in its 
semiclassical limit. The puzzle that poses itself is, how a background 
independent theory can describe a background dependent theory such as
QFT on CS {\it all}. This question is the central topic of the present 
paper.   

We still do not have a working theory of quantum gravity, however, there 
are several proposals which are still under construction. One of them,
Loop Quantum Gravity (LQG) {\cite{rovellibook,thiemannbook}} indeed 
is a 
background independent approach to the quantisation of general 
relativity and in the 
last 20 years a lot of effort has been put in the rigorous definition of 
the mathematical framework of this theory. Background independence is a 
feature that distinguishes LQG from most other quantum field theories 
constructed so far,  as for example the standard model of elementary 
particle physics. When examining ordinary quantum field theory with care 
one realises that much of the structure we are  used to from the 
particle physics point of view, including { for instance} the 
construction of 
a Fock space as the fundamental Hilbert space of the theory, the concept 
of a particle as a certain mode--excitation of the field, the notion of 
causality and { the notion of a vacuum state}, are strongly linked to 
the presence of a given background spacetime. For a quantum theory of 
general relativity there is no such background spacetime available 
because we are quantising the dynamics of { spacetime} itself, and thus 
it should not be too surprising that the fundamental structures 
appearing in a quantisation of general relativity differ a lot from the 
ones used in ordinary quantum field theory.\\
{The natural Hilbert space} to describe the quantum dynamics of LQG is 
not a Fock space as one might expect -- in fact the Hilbert $\HC_{\rm 
AIL}$ space used in LQG, named the Ashtekar--Isham--Lewandowski Hilbert 
space \cite{AIL,AL}, 
is the space of square--integrable functions over a set of appropriately 
generalised $SU(2)$--connections on a 3 dimensional manifold and does 
not bear a lot of similarity with the Hilbert spaces known from ordinary 
QFT. However, when trying to find irreducible representations of the { 
kinematical algebra} of GR, { also called the holonomy -- flux -- 
algebra in the LQG context}, one is naturally led to this space. { 
Analogous to the well known Stone--von Neumann theorem in quantum 
mechanics, there exists even an
uniqueness result \cite{LOST, Fleischhack} }, that states that under 
certain assumptions ({ spatial} diffeomorphism invariance being the main 
one) the representation used in LQG { associated with} $\HC_{\rm AL}$ is 
unique.
 The structure of this Hilbert space is quite different from the ones 
known from ordinary quantum field theory on a given background: It is 
manifestly background independent and its elements can be written in 
terms of so called spin network functions, functions living on certain 
embedded graphs which are labeled by { irreducible} representations of 
$SU(2)$. Furthermore, the flux operators,{ corresponding to the momentum 
operators in the canonical picture}, encode geometrical information 
about the quantum state turn out to fulfil a non--commutative 
algebra.\\
The relation between LQG and classical general relativity has been 
studied in detail \cite{GCSI, GCSII, GCSIII, GCSIV, AQGII} and it turns 
out that LQG, or rather its recently proposed modification coined 
algebraic 
quantum gravity, see \cite{AQGI, AQGII, AQGIII}), has the right 
semiclassical limit { at least as far as the infinitesimal dynamics is 
concerned.}
\\
However, if LQG is a fundamental quantum theory for gravity and  matter, 
then it { should} be possible to { rediscover} ordinary quantum field 
theory on a fixed background { within a sector} of the fully background 
independent framework of LQG. There has been some work into this 
direction \cite{sahlmann_thiemann_curved1, sahlmann_thiemann_curved2}, 
but some issues remained unresolved: First, one did quantise at the 
level of the kinematical Hilbert space. Therefore all operators in 
question were not gauge invariant, the coherent states that one was 
using were not gauge invariant and thus one 
might rightfully ask 
whether the constructions displayed there would survive upon passage to 
the physical Hilbert space. Next, the matter Hamiltonians used there 
were just the matter contributions to the Hamiltonian constraint and one
did not have a convincing argument why those should be used as 
Hamiltonians driving the physical time evolution of the system. 
Finally, backreaction effects were not taken into account.

The aim of 
this article is to try to improve the understanding 
of how a quantum field theory { of matter} on a given background can 
emerge out of a theory of quantum gravity. In order to deal with 
the afore mentioned constraints we  
carry out a reduced 
phase space 
quantisation 
of general relativity plus matter degrees of freedom using a dynamical 
reference frame given by some dust fields using methods from 
\cite{ghtw1, ghtw2, AQGIV}. Thus we arrive at a true (i.e. 
non--vanishing) physical Hamiltonian that generates time evolution in 
the reference frame described by these dust fields. 
{ Since in contrast to the total Hamiltonian constraint the physical 
Hamiltonian is non -- vanishing, in general, the physical Hamiltonian 
operator will involve more complicated  terms than just the  total 
Hamiltonian constraint. In particular  it involves a square root of 
polynomials of the contributions to the constraints associated with the 
non -- dust degrees of freedom. To deal with this square root is a 
difficult technical subject of its own which is not the main concern of 
this paper. Thus, we consider a regime of the theory for which the 
Taylor expansion of the square root can be restricted to the dominant 
term which is just the integrated non -- dust contributions to the
Hamiltonian constraint.} Therefore we get an a posteriori justification 
for why one can build a matter QFT just using the matter parts of the 
Hamiltonian constraints as dynamical operators which otherwise would 
seem rather ad hoc. Of course, there is also the gravitational 
contribution which we will take care of in this paper. That this 
is possible relies on the judicious choice of the Brown -- Kucha{\v r}
dust fields as rods and clocks.\\
When examining LQG and ordinary QFT from a mathematical point of view it 
turns out that there are two main issues which need to be clarified when 
the latter should emerge out of the former in some appropriate limit: 
First, although LQG is a true continuum theory, its basic building 
blocks are defined on graph like structures. On the other hand, the Fock 
space of ordinary QFT uses the continuum properties of the underlying 
spacetime and one has to understand how one can arrive at such a space 
starting with a discrete theory at least in some limit. 
\\
Second, the flux operators in LQG, which are the basic building blocks 
if we were to construct a `metric operator' as a quantum analog of the 
classical metric form a non--abelian operator algebra in LQG. This 
non--commutativity originates in the fact that the fluxes are obtained 
by smearing the sensitised triads along hypersurfaces of codimension one 
(and thus, singular ones from the three dimensional point of view) 
\cite{ashtekar_corichi_zapata} which causes the associated Hamiltonian 
vector fields to be non commutative. Hence in any representation of this 
algebra on a Hilbert space, the non commutativity prohibits that the 
three geometry operators which couple to the matter fields can be 
simultaneously diagonalised. This implies an obstacle to directly 
applying the framework of QFT on CS when choosing a representation 
for the matter fields because there one relies on a commutative 
background spacetime. While much work on QFT on non commutative 
spacetimes has been conducted (see e.g. 
\cite{Dopplicher,Dopplicher1,Bahns,Bahns1,Bahns2} and 
references therein), there one assumes a non commutative structure
in the coordinates rather than the metric and therefore we are not in 
the position to apply that framework. Thus, if one wishes to import QFT 
on CS techniques into LQG, one first has to understand how to 
``Abelianise'' the metric operators in appropriate regime.\\
\\
In this article we will focus on the first point and leave the latter 
open for future research: Namely we will construct a quantum theory for 
gravity plus matter that is close to LQG in the sense that it is defined 
on a fundamentally discrete entities (algebraic graphs), but the 
geometrical operators are defined as ordinary (commuting) multiplication 
operators. This is possible by choosing an algebra different 
from the holonomy -- flux algebra which 
shares some of its features (namely the discreteness of its fundamental 
building blocks) but deviates from it in other essential points (mainly 
the non--commutativity of geometrical operators is avoided). We choose a 
representation of this algebra which is very close to a (background 
independent) Fock representation\footnote{Essentially, we smear the 
continuum co-triad along paths and the conjugate momentum along 
surfaces, similar as connections and densitised triads in LQG. 
Notice that the smeared co-triads in contrast to the holonomy of a 
connection do not transform locally under gauge transformations 
which makes it difficult to construct gauge invariant objects from 
them, in contrast to the Wilson loops. However, in this paper we  
consider a top to bottom approach and define an abstract algebra on an
algebraic graph which approximates the continuum algebra and which 
transforms locally under gauge transformations.} and therefore the 
discreteness of the spectrum of geometrical operators of LQG is not true 
in this representation\footnote{There is no contradiction to the 
uniqueness theorem because we use a different algebra.}  
Using this algebra we 
examine how a Fock space for the matter degrees of freedom on a fixed 
background can emerge out of a full quantum theory for gravity plus 
matter. 

The idea then for how to import QFT on CS techniques into quantum 
gravity is to employ a Born--Oppenheimer Decomposition (BOD) of the 
Hilbert space: In the chosen representation for the gravitational field,
the wave functions depend on the co-triad configuration $e$. For each co 
-- 
triad configuration we consider a matter Hilbert space ${\cal 
H}_e$ labelled by the 
co -- triad. That matter Hilbert space is naturally chosen as a Fock 
space defined by the natural creation and annihilation operators on the 
formally ultrastatic\footnote{A spacetime metric is called ultrastatic 
if it is static (i.e. stationary and the shift vanishes) and the lapse 
equals unity. Thus the non -- trivial components of the four -- metric 
are encoded in the three -- metric which is not explicitly time 
dependent. There is
no explicitly time dependence here because our Hamiltonian is not 
explicitly 
time dependent by virtue of the choice of the material reference frame.} 
spacetime defined by the co -- triad, whenever it 
is non -- degenerate. When the co -- triad is degenerate, the QFT on CS
construction does not work and in fact the Hamiltonian operator 
is ill -- defined in this case. Its domain of definition therefore must 
be chosen as to exclude this case.  
The fibre Hilbert 
spaces ${\cal H}_e$ are then glued together in a direct integral with 
respect to the spectral measure defined by the projection 
valued measures of the commutative algebra of 
co -- triad operators. The cotriad operator and the momentum 
operators respectively act by multiplication and (suitable) derivation 
with 
respect to $e$ in the 
fibres respectively. Also the matter field 
operators, being expressible as linear combinations (with $e$ dependent 
``coefficients'') of creation and annihilation operators, preserve 
these.
This works when the Hilbert space for the total 
system is separable. Conversely, choosing a spectral measure and an
action of the momenta conjugate to the triads with 
respect to which they are self -- adjoint defines a separable Hilbert 
space representation. Notice that in this construction the question of 
unitary equivalence of the fibres ${\cal H}_e$ never arises (they will 
typically be unitarily inequivalent). 
We will 
generalise this construction and start with a non -- separable Hilbert 
space based on von Neumann's Infinite Tensor Product. Now the 
separable Hilbert spaces ${\cal H}_e$ (and uncountably infinitely 
many more) arise as strong equivalence class separable sectors in the 
matter ITP labelled by the corresponding vacua $\Omega_e$. They are 
automatically orthogonal in the ITP inner product when the $\Omega_e$ 
lie in different strong equivalence classes. The advantage of the ITP 
over the separable Hilbert space construction is that we do not need to 
choose a spectral measure (and the ${\cal H}_e$ for degenerate ${\cal 
H}_e$ which could be defined e.g. by an $e$ independent Fock 
representation) and 
that it is easier 
to find a self -- adjoint action of the 
operator conjugate to the triad. Moreover, the ITP contains vectors of
the form $\Omega_g\otimes \Omega_m$ where $\Omega_g,\Omega_m$ lie in the 
gravitational and matter ITP respectively. This not possible in the 
separable construction (at least not for non compact spacetimes 
for which an infinite algebraic graph is essential) because this would 
mean that 
the same 
$\Omega_m$ lies in all the ${\cal H}_e$ which is a contradiction when 
the ${\cal H}_e$ are not all unitarily equivalent.

The use of Born--Oppenheimer Decomposition in the context of quantum 
gravity 
has a long tradition, see for example \cite{kiefer_book} 
and references therein for a historical review of the 
Born--Oppenheimer 
method in quantum geometrodynamics or \cite{rovelli_vidotto} for a 
recent application in the context of spinfoam models. However, to the 
best of our knowledge, the connection between the BOD and QFT on CS
has not been much exploited yet. As already mentioned, one would like to 
use the BOD also for the representation underlying LQG, however, at the 
moment there seems to be no promising idea for how to do this when 
those slow degrees of freedom that couple to the fast degrees of freedom 
cannot be simultaneously diagonalised.  
\vspace{.2cm}\\
The structure of this article is as follows:\\ In section 
\ref{sec:classical} we very briefly review how one can construct a true 
physical Hamiltonian for general relativity plus arbitrary matter at the 
classical level using a preferred reference frame given by four dust 
fields. For details concerning this construction see \cite{brown_kuchar, 
ghtw1, ghtw2}. Then we define a regime for which it is justified 
to approximate the physical Hamiltonian  
by just the { integrated contributions of the non -- dust 
degrees of freedom to the Hamiltonian constraint}. In the remainder of the 
article we will use this 
approximation to define dynamics in the dust frame. { Next,} we rewrite 
classical GR in terms of an $SU(2)$--cotriad and its canonically 
conjugate momentum which is essentially a densitised extrinsic 
curvature. The smearing of these objects along one dimensional paths and 
two dimensional surfaces respectively defines a new Poisson algebra for 
classical GR which separates the points of the phase space and which we 
will call the {\it (extrinsic) curvature -- length algebra} because the 
square root of the trace of the square of the cotriad along short paths 
is approximately the length along that path\footnote{It is 
straightforward to construct a representation of this algebra in the 
continuum very close 
to the one of LQG by taking as configuration variables the exponentials 
of the cotriads (one for each direction in the Lie algebra) along paths 
which would substitute the holonomies of LQG and then one would use 
the Ashtekar -- Lewandowski measure for $U(1)^3$. However, the 
construction of gauge invariant functions in the continuum from these
Abelian holonomies works only in the limit of short paths which is 
a limit too singular in this representation because it is not strongly 
continuous under path deformations. This is why we proceed to the 
algebraic formulation here.}.    
In section \ref{sec:quantum} we construct a 
representation of the curvature -- length algebra and of a  
scalar field of 
Klein--Gordon type\footnote{A generalisation to gauge fields or fermions 
is straightforward but we will only consider a scalar field for 
conceptual clarity.} on an abstract 
algebraic graph in the spirit of \cite{AQGI, AQGII, AQGIII, AQGIV} and 
the essential feature which distinguishes this theory from { ordinary} 
LQG is that { cotriads are quantised as multiplication operators} which 
will be essential for the semiclassical analysis in the second half of 
this article.\\
In section \ref{sec:classical_limit} we construct semiclassical states 
for the gravitational sector of this quantum theory.\\
In section \ref{sec:analysis} we introduce the Born -- Oppenheimer 
Decomposition for our system and the associated approximation scheme 
for the full spectral problem, which still takes backreaction effects 
into account. Here the gravitational variables are 
treated as the `heavy' ones and the matter variables as the `light' 
ones. Such an approximation scheme is justified in regimes where a 
description in terms of interacting quantum fields on an almost 
classical background spacetime makes sense because for time scales on 
which particle interactions take place the change in geometry are truly 
small. We illustrate how this works using a simple 
minisuperspace model. In the appendix we list some facts about the ITP 
for the benefit of the reader.

\section{Classical theory} \label{sec:classical}

\subsection{Brown--Kuch\v{a}r dust reduction and observables for General 
Relativity} \label{sec: brown_kuchar}

The problem of observables, that is, calculating quantities which are 
invariant under the gauge group of the theory is much more severe in 
general relativity compared to other gauge theories such as 
electrodynamics or $SU(N)$ Yang--Mills theories. The reason is twofold: 
First, in general relativity the gauge group is given by Diff$(M)$, the 
diffeomorphism group of the four dimensional underlying manifold. This 
leads to a rather complicated constraint algebra (often called the {\it 
hypersurface deformation algebra} or {\it Dirac algebra} in the 
literature) which is neither finite dimensional nor an honest Lie 
algebra (it has phase space dependent structure functions and not 
structure constants). Second, the issue of observables is strongly 
related to the background independent nature of the theory: Taking the 
manifold $M$ just as an auxiliary structure, it is not meaningful to 
talk about fields evaluated at some point on this manifold. The true 
gauge dependent degrees of freedom should be defined as {\it relations 
between dynamical fields}.\\
Such relational observables were introduced first in \cite{ProbTime1, 
ProbTime2, bergmann, bergmann_komar} and later refined in
\cite{rovelli1, rovelli2, GaugeUnfixing, dittrich1, dittrich2}. 
In retrospect it turns out that there is a one to one correspondence 
between a choice of gauge fixing conditions with an associated reduced 
phase space and a choice of relational observables. The correspondence 
is established by choosing as rods and clocks of the relational 
observables precisely the gauge fixing conditions of the reduced 
framework, see e.g. \cite{ghtw1,ht1}. The explicit expression 
of these observables in terms of the non gauge invariant degrees of 
freedom is rather involved, see e.g. \cite{dittrich_tambo1, 
dittrich_tambo2} for a perturbative treatment. Fortunately, it is not 
needed for what matters is the Poisson algebra between the 
observables (i.e. true degrees of freedom) and the associated 
physical Hamiltonian (i.e. reduced Hamiltonian) \cite{thiemann1}. The 
reduced phase space for vacuum general relativity has a very complicated 
symplectic structure because the rods and clocks that one can 
construct necessarily lead to spatially non -- local objects. However, 
in the presence of suitable matter these complications can be avoided. 
Here by suitable we mean matter that comes as close as possible to 
the mathematical idealisation of a test observer fluid 
and such that the physical Hamiltonian that drives the 
physical time evolution of the observables comes as 
close as possible to the standard model Hamiltonian \cite{thiemann2}
when the spacetime is close to flat.
In \cite{ghtw1, ghtw2} such a suitable matter system was identified 
as the pressureless dust matter fields introduced by Brown and 
Kuch\v{a}r in 
\cite{brown_kuchar}. One considers the enlarged system 
$S_{\rm EH} +  S_{\rm matter}+ S_{\rm dust}$ where $S_{\rm EH}$ is the 
usual Einstein--Hilbert action, $S_{\rm matter}$ includes all possible 
other standard matter that one likes to couple to gravity and the dust 
action  $S_{\rm dust}$ is given by
\be
S_{\rm dust} = -\frac{1}{2}\int\limits_{M}d^4 X \sqrt{|\det(g)|}\rho 
[g^{\mu \nu} U_\mu U_\nu + 1] \quad  \label{dustaction}
\ee
Here $M$ is the four dimensional (spacetime) manifold which can 
topologically be identified with $\mathbb{R} \times \cal{X}$ for some 
three dimensional manifold $\cal{X}$ of arbitrary topology and $X \in M$ 
are local coordinates on $M$. $g_{\mu \nu}(X)$  with $\mu, \nu = 
0,1,2,3$ denotes a (Pseudo--) Riemannian metric on $M$. $U \in T^*M$ is 
a one form defined as $U = -dT + W_jdS^j$, with $j 
=1,2,3$, for some scalar fields $T, W_j, S^j \in C^\infty(M)$. So 
finally the action (\ref{dustaction}) is a functional of $g_{\mu \nu}$ 
and eight scalar fields $\rho, T, W_i, S^i$.\\
The coupled system $S_{\rm EH} + S_{\rm matter} + S_{\rm dust}$ seems to 
be very complicated at first sight, and indeed when performing a 
Hamiltonian analysis one realises that it is a second class constrained 
system. However, the second class constraints can be solved and on the 
reduced phase space the Dirac bracket is identical to the usual Poisson 
bracket. Moreover, if one chooses $T, S^j$ as the four clock fields then 
it turns out that one can explicitly construct observables associated 
to the spatial three metric $q_{ab}(x)$ and their conjugate momenta 
$p^{ab}(x)$ in the dynamical reference frame $(\tau, \sigma^a)$ defined 
through the readings of the dust fields $T, S^j$. We will not give the 
details of this construction here and the interested reader should 
consult \cite{ghtw1, ghtw2} for all the details. What will be important 
for the purpose of this article is that we can construct a {\it physical 
Hamiltonian} $H_{\rm phys}$ that generates {\it physical equations of 
motion} for the {\it gauge invariant}\footnote{We will use capital 
letters to denote the gauge invariant observables in the reference frame 
given by the dust clocks $T, S^j$. Thus, by $Q^{ab}$ we denote the gauge 
invariant three metric associated to the gauge variant 3--metric 
$q_{ab}$ and by $P^{ab}$ the gauge invariant canonical momenta 
associated to the gauge variant momenta $p_{ab}$. For the exact 
construction of these gauge invariant quantities see the articles cited 
above.} 3--metrics $Q_{ab}(\sigma)$ and their respective momenta 
$P^{ab}(\sigma)$. These observables are functions on the ``dust 
manifold'' $\Sigma$, i.e. $\sigma \in \Sigma$ does not label a point in 
${\cal X}$ but a point in the range of the clock fields $S^j$. The 
explicit form of the physical Hamiltonian reads
\be
\label{Hphys}
H_{\rm phys} := \int\limits_{\Sigma}d\sigma \sqrt{C^2 - Q^{ab} C_aC_b} =  
\int\limits_{\Sigma}d\sigma H(\sigma)
\ee
where $C:=C^g(Q,P)+ C^m(Q,\Pi,\Phi)$ and $C_a:=C_a^g(Q,P) + 
C^m_a(Q,\Pi,\Phi)$ are the contributions to the Hamiltonian and 
diffeomorphism constraints of 
general relativity and the matter sector written in terms of gauge 
invariant quantities $Q_{ab}(\sigma), P^{ab}(\sigma)$ and the 
corresponding matter observables denoted by 
$\Pi(\sigma),\Phi(\sigma)$. Also we introduced the Hamiltonian density 
denoted by $H(\sigma)$. 

Note that $C$ and $C_a$ are non vanishing because only the total 
Hamiltonian and diffeomorphism constraint including the contribution of 
the dust vanishes and the energy momentum density of the dust, which are 
constants of the physical motion, 
must be 
non vanishing anywhere in order that it provides a bona fide system of 
coordinates.
Note 
that the matter degrees of freedom should be just understood 
symbolically since we have not yet decided which matter components we 
actually want to consider. In general the algebra of observables 
involves a Dirac bracket rather than the Poisson bracket. However, for 
this particular  choice of clock variables, $Q_{ab}$ and $P^{ab}$ as 
well as the corresponding matter observables form mutually commuting  
canonical pairs, that is their Poisson bracket is given by $\{ 
Q_{ab}(\sigma) , P^{cd}(\sigma')\}= \delta_{(a}^c 
\delta_{b)}^d\delta(\sigma, \sigma')$ and one can show that the physical 
equations of motion are given by
\be
\dot{Q}_{ab} = \{ H_{\rm phys}, Q_{ab} \}, \quad 
\dot{P}^{ab} = \{ H_{\rm phys}, P^{ab} \}
\ee
and likewise for the matter sector.  \\
\\
The consistency of the dust picture with the usual gauge variant 
description of black hole and cosmological spacetimes has been verified 
in \cite{ghtw1, ghtw2,gtt}.


\subsection{Approximation for the physical Hamiltonian}
\label{sec:approx}
The physical Hamiltonian shown in equation (\ref{Hphys}) includes a 
square root under the integral. This square root originates from solving 
the Hamiltonian constraint for GR coupled to dust and possibly other 
standard matter for the dust momentum P, the momentum conjugate to the 
dust time $T$. We want to discuss how the reduced framework when 
quantised is related to quantum field theory on curved {{spacetimes.} In 
contrast to the standard framework we will also quantise the
gravitational sector of the theory by quantising $H_{\rm phys}$ which 
contains all (gauge -invariant) degrees of freedom except the dust ones. 
Although the physical Hamiltonian $H_{\rm phys}$ has been quantised in 
its full form using Loop Quantum Gravity techniques and 
spectral theory \cite{AQGIV}, in 
this paper we will consider $H_{\rm phys}$ in a certain approximation 
which we will now discuss.

First of all, the derivation of $H_{\rm phys}$ in \cite{ghtw1} shows 
that one has the two anholonomic constraints $C^2-Q^{ab} C_a C_b\ge 0$
and $C>0$  
on 
the constraint surface as otherwise the momentum $P$ conjugate to $T$ 
becomes imaginary and because one was using phantom dust rather than 
dust. This is why the argument of the square root is 
classically forced to be non negative. Thus, classically we can write 
$H=C\sqrt{1-\frac{Q^{ab} C_a C_b}{C^2}}$ and know that the Taylor 
expansion of the square root 
\be
\label{Hdapprox}
H(\sigma)
&=& C\sqrt{1 - \frac{Q^{ab} C_aC_b}{C^2}}(\sigma)\nonumber\\
&\simeq& C  \left(1 - \frac{1}{2} \frac{Q^{ab} C_aC_b}{C^2} + 
\frac{1}{8}\left[\frac{Q^{ab} C_aC_b}{C^2}\right]^2\right)(\sigma) 
+\OC([Q^{ab}C_aC_b/C^2]^3)
\ee
converges. One could now argue that square root and its Taylor 
expansion 
(\ref{Hdapprox}) provide 
classically
equivalent starting points and therefore instead of defining the square 
root via the spectral theorem, one could quantise the Taylor expansion 
directly. To do the latter, one would need to order the terms in each 
order of the expansion symmetrically and one would need to define the 
operator corresponding to $1/C$ on a dense domain where the expansion is 
valid. To that end, notice the identity
\be \label{identity}
\frac{1}{C}
=\lim_{\epsilon\to 0+} \frac{C}{C^2+\epsilon^2}
=\lim_{\epsilon\to 0+} \frac{1}{2} 
[\frac{1}{C+i\epsilon}+\frac{1}{C-i\epsilon}]
=\lim_{\epsilon\to 0+} \frac{1}{2} 
[R(i\epsilon)+R(-i\epsilon)]
\ee
where $R(z)=(C+z)^{-1}$ denotes the resolvent of $C$ at $z\not\in {\rm 
spec}(C)$. Thus if we can quantise (the smeared) density 
$C(\sigma)$ as a self -- adjoint operator, then the resolvent 
$R(\pm i\epsilon)$ is a bounded operator at finite $\epsilon$. The 
removal of the regulator will of course only be possible on a 
suitable dense domain of the Hilbert space, for instance on analytic 
vectors for $C$. 

In order not to deviate from the main thrust of the paper, we 
will neglect the higher order contributions in (\ref{Hdapprox}) by 
considering a subspace of the Hilbert space for which in the sense 
of expectation values $<Q^{ab}C_aC_b> \ll <C>^2$. We will come back 
to the higher order terms in a future publication. Thus for the 
remainder of the paper we will assume that 
\be
H_{\rm phys}&\simeq&  \int\limits_{\Sigma}d\sigma  C(\sigma)
\ee
is a good approximation. For instance, for spacetimes close to a 
homogeneous one the approximation should be very good as one 
can see by expanding in terms of the number of spatial 
derivatives. Then the zeroth order contribution 
to the total diffeomorphism 
constraint vanishes identically so that $H_{\rm phys}=\sqrt{C^2}=C$ is 
true without any approximation in zeroth order. To define the 
sectors of validity of 
the approximation in the quantum theory is more delicate but one could 
use semiclassical perturbation theory \cite{aqg2}. See also 
\cite{AshtekarKaminskiLew} for related ideas.

\subsection{Canonical transformation} \label{sec:canonicaltrafo}

Using the gauge fixed formalism described above, where one has a 
preferred reference frame given by the Brown--Kuch\v{a}r dust clocks,
general relativity in terms of observables can be defined as follows: 
Let $\mathcal{M}$ be an (infinite dimensional) symplectic manifold and 
$(Q_{ab}(\sigma), P^{ab}(\sigma))$ a coordinate basis of $\mathcal{M}$. 
$Q_{ab}(\sigma)$ can be interpreted as a (physical, i.e. gauge 
independent) Riemannian metric\footnote{This identification can be made 
precise by demanding appropriate boundary conditions for $Q_{ab}, 
P^{ab}$ and specifying the function space in which they live. See 
\cite{ghtw1} for more details and appropriate boundary conditions.} on a 
3--dimensional manifold ${\Sigma}$ with local coordinate system 
$\{\sigma^a \}_{a=1}^{3}$ defined through the readings of the dust 
fields and $P^{ab}(\sigma)$ is related to the extrinsic curvature of 
${\Sigma}$ in a 4 -- dimensional {{globally hyperbolic} manifold $M$ 
which in turn is isomorphic to $\mathbb{R}\times {\Sigma}$. ${\Sigma}$ 
as a Riemannian manifold carries geometric structures such as a unique 
{{torsionfree}  covariant derivative $D$ or the Ricci scalar $R(Q)$. 
$Q_{ab}(\sigma)$ and $P^{ab}(\sigma)$ form a canonical pair, i.e. the 
symplectic 2--form $\Omega \in T^*\mathcal{M} \times T^*\mathcal{M}$ is 
given by $\Omega := \int d\sigma dP^{ab}(\sigma)\wedge dQ_{ab}(\sigma)$. 
Furthermore let
\be
C^{g} & := & \frac{\kappa}{\sqrt{\det Q}} \big[ Q_{ab} Q_{cd} - 
\frac{1}{2} Q_{ac} Q_{bd}  \big] P^{ac} P^{bd} - 
\frac{1}{\kappa}\sqrt{\det Q} R(Q)\\
C_a^g & := & -2 Q_{ab} D_c P^{bc}
\ee
be $C^\infty$ functions on $\mathcal{M}$, $X_{C^g}$, $X_{C^g_a}$ their 
respective Hamiltonian vector fields and $\kappa = 8 \pi G/c^3$ the 
gravitational coupling constant. These functions are the gravitational 
parts of the {\it Hamiltonian} and {\it diffeomorphism constraint} 
respectively.\\
If we would work in the standard formalism of GR where $q_{ab}, p^{ab}$ 
are the gauge dependent 3--metric and its canonical momentum then $C^g$ 
and $C^g_a$  (after the replacement $Q_{ab} \rightarrow q_{ab}, P^{ab} 
\rightarrow p^{ab}$) would generate gauge transformations and only 
quantities which are constants along the flow lines of their respective 
Hamiltonian vector fields would have a physical interpretation. However, 
in the enlarged phase space described in the last section, $C^g$ and 
$C^g_a$ are just the {\it gravitational parts} of the constraints and, 
without taking into account the parts coming from the dust fields, do 
not generate gauge transformations. Due to the gauge fixing and 
construction of manifestly gauge invariant quantities $Q_{ab}(\sigma), 
P^{ab}(\sigma)$ we have a physical Hamiltonian that generates time 
evolution in the dust--frame, see equation (\ref{Hphys}).
For the purpose of what follows we introduce an orthonormal frame at 
each point $\sigma \in {\Sigma}$ and define the cotriad\footnote{Again, 
we use a capital letter $E_a^i$ to make clear that this is a gauge 
invariant version of the cotriad. It should not be mistaken for the 
densitised triad used in LQG.} $E_{a}^i$ through $Q_{ab}(\sigma) = 
\delta_{ij} E_{a}^i E_{b}^j$. It can easily be seen that this definition 
is invariant under $SO(3)$ or more general  $SU(2)$ rotations, thus the 
cotriads can be interpreted as $su(2)$--valued one forms with 
$i,j,k,\dots = 1,2,3$ $su(2)$--indices. Furthermore define the triad 
$E^a_i$ to be the inverse of the cotriad through $E^a_i E^i_b = 
\delta_a^b, E^a_i E_a^j = \delta_i^j$.\\
Now we perform the following coordinate transformation:
\be
(Q_{ab}, P^{ab}) & \rightarrow & (E_a^j, P^a_j  :=  2 P^{ab}  E_b^j) 
\quad.
\ee
One can easily check that this transformation is indeed a canonical one, 
i.e. it leaves the symplectic 2--form invariant and thus $(E_a^i, 
P^a_i)$ form a canonical pair
\be
\{ P^a_k(\sigma), E^i_b(\sigma')  \} & = & \kappa \delta_b^a \delta_k^i 
\delta(\sigma, \sigma') \quad .
\ee
However, the phase space spanned by $(P^a_k, E_a^k)$ is larger than the 
one spanned by $(P^{ab}, Q_{ab})$: Only the symmetrised product $P^{ab} 
= \frac{1}{2}P^{(a}_j E^{b)}_j :=\frac{1}{4}(P^a_jE^b_j + P^b_jE^a_j) $ 
survives when we invert the canonical transformation and want to 
{{transform} back to ADM variables, which means that the canonical 
transformation $(Q_{ab}, P^{ab})  \rightarrow  (E_a^j, P^a_j)$ is not 
one--to--one but one--to--many. To get rid of the additional degrees of 
freedom we must demand $P^{[a}_j E^{b]}_j := \frac{1}{2}(P^a_j E^b_j - 
P^b_j E^a_j) = 0$ which after some algebraic manipulations leads to a 
constraint
\be
G_i := \epsilon_{ijk}P^a_j E_a^k \quad ,
\ee 
the so called SU(2)--Gauss constraint\footnote{In analogy with the 
formulation of GR in terms of Ashtekar's variables we still call this 
the Gauss--constraint although it does not have the form known from 
Yang--Mills theories.}.\\
Modulo Gauss constraint the gravitational parts of the Hamiltonian and 
diffeomorphism constraints can be written as $C^g = \frac{\kappa}{4 \det 
(E)} [E_a^i E_b^i \delta_m^k - \frac{1}{2}E_a^k E_b^m] P^a_k P^b_m - 
\frac{\det (E)}{\kappa} R(E)$ and $C^g_a = - E_a^i D_b(e) P^b_i$ 
respectively where $R(E)$ denotes the Ricci scalar written in terms of 
the cotriad and $D(E)$ is the covariant derivative compatible with the 
cotriad.\\
So finally we arrive at a classically equivalent formulation of general 
relativity in the dust--frame, which we want to summarise in the 
following paragraph:\\
The phase space $\mathcal{M}'$ is spanned by an $su(2)$--valued 
one--form $E_a^{i}$ and its canonically conjugate momentum $P^a_j$, a 
vector density which takes values in $su(2)$ as well. The phase space is 
subject to the Gauss constraint
\be
G_i & = & \epsilon_{ijk} P_j^a E_a^k \quad ,
\ee
which means that only quantities which are invariant along the flow 
lines of its associated Hamiltonian vector field have a physical 
interpretation. In the gauge fixed formalism we have a physical 
Hamiltonian
\be
H_{\rm phys} = \int\limits_{\Sigma}d\sigma\sqrt{(C^g+C^m)^2 - 
Q^{ab}(C^g_a+C_a^m)(C^g_b+C_b^m)}
\ee
that generates motion in dust--time and in our variables the 
gravitational parts of the Hamiltonian and diffeomorphism constraints 
are given by
\be
C^g & = & \kappa G_{ab}^{ij} P^a_i P^b_j - \frac{1}{\kappa}\det (E) R(E) 
\label{C^g} \\
C^g_a & = & -E_a^i D_b(E)P^b_i \quad,
\ee 
As before $C^m$ and $C^m_a$  denote all possible standard matter 
contributions that one might couple to gravity and we defined the 
deWitt--type supermetric $G_{ab}^{ij} := \frac{1}{4 \det (E)} [E_a^k 
E_b^k \delta^{ij} - \frac{1}{2} E_a^i E_b^j]$.
Thus, in contrast to the formulation of general relativity in terms of 
Ashtekar's variables the Hamiltonian constraint looks similar to a 
standard Hamiltonian composed out of of a kinetic term $\propto P^2$ and 
a potential $\det (E) R(E)$. 

The choice of these ``hybrid variables'' 
lies in between those of connection dynamics and geometrodynamics. 
Geometrodynamics is not sufficient if we are interested in coupling 
fermionic matter. Furthermore, it is much harder to smear symmetric 
tensors over submanifolds in a spatially diffeomorphism covariant 
manner than it is possible for the one and two forms $e,\;\ast P$ 
respectively. The drawback of using cotriads rather than connections, 
as already mentioned in the introduction, is that the holonomy 
transforms locally under SU(2) gauge transformations while the integral 
of the cotriad along one dimensional curves does not. However, in the 
algebraic setting that we choose in the next section, this will not pose 
any problem. 
\vspace{.2cm}\\
For what follows we need the explicit expression of the curvature 
$R(E)$--term in the Hamiltonian constraint. The cotriad $E_a^i$ together 
with the compatibility condition $D_a E_b^i \equ 0$ uniquely defines a 
spin connection $\Gamma_a^i$ through $D_a E_b^i = : \partial_a E_b^i + 
\Gamma_{ab}^c E_c^i + \epsilon_{ijk} \Gamma_a^j E_b^k$ where 
$\Gamma_{ab}^c$ are the components of the Levi--Civita connection. In 
terms of the cotriad and its inverse this spin connection can be written 
as
\be
\Gamma_a^i = - \frac{1}{2}\epsilon_{ijk}E_j^b (E^k_{a,b} - E^k_{b,a} + 
E^c_k E^m_a E^m_{c,b}) \quad ,
\ee
where we used $E^i_{a,b}$ as a short hand notation for $\partial_b 
E^i_a$.\\
Using the above covariant derivative one can define a curvature tensor 
$R_{ab}^j$ in the obvious way as\footnote{This definition is only 
meaningful if $\dim({\Sigma}) = 3$ where the fundamental and the adjoint 
representation of the gauge group have the same dimension.}
\be
[ D_a, D_b] v_j =: \epsilon_{jkl}R_{ab}^kv_l
\ee
for every $su(2)$--valued scalar function $v_j$ on $\Sigma$. In terms of 
the spin connection this curvature tensor can be given explicitly as
\be
R_{ab}^l = 2 \partial_{[a}\Gamma_{b]}^l + \epsilon_{lkj}\Gamma_a^k 
\Gamma_b^j \quad .
\ee
Finally the Ricci scalar in terms of cotriads can be written as
\be
R(E) & = & \epsilon_{jkl}R_{ab}^l E_j^a E_k^b = 2 \epsilon_{jkl} E_j^a 
E_k^b \partial_a \Gamma_b^l + (E_k^a \Gamma_a^j)^2 - E_j^a E_k^b 
\Gamma_a^k \Gamma_b^j \quad .
\ee
In this form the Ricci scalar is a function of the cotriad $E_a^i$ and 
its inverse $E^a_i$. Having a quantised version of this theory in mind, 
it will turn out to be more convenient to rewrite $R$ in terms of the 
cotriad and its determinant only which can be achieved by employing the 
identity $E^a_i = \frac{1}{2 \det (E)} \epsilon^{abc}\epsilon_{ijk} 
E_b^j E_c^k$. Thus the 3 terms in the preceding expression take the form
\be
(E_i^a \Gamma_a^i)^2 & = & \frac{1}{4 (\det (E))^2} 
\epsilon^{abc}\epsilon^{def}E_a^k E_d^l E_{b,c}^k E_{e,f}^l  
\label{C^g_1} \\
-E_i^a E_j^b \Gamma_a^j \Gamma_b^i & = & -\frac{1}{4 (\det 
(E))^2}\epsilon^{abc}\epsilon^{def}\big[ E^i_c(E^i_{d,b} - E^i_{b,d}) + 
E^j_d E^j_{c,b}  \big]   \big[ E^k_f (E^k_{a,e} - E^k_{e,a}) + E^l_a 
E^l_{f,e}  \big] \label{C^g_2}	\\
2\epsilon_{ijk} E^a_i E^b_j \partial_a \Gamma_b^k & = & -\frac{1}{ (\det 
(E))^2}\epsilon^{abc}\epsilon^{def} E^i_c E^i_f \big[ E^j_e E_{b, da}^j 
+ E^j_{b,a}E^j_{e,d} + E_b^j E^j_{e, da}  \big] \nn \\
& & + \frac{1}{2 (\det 
(E))^3}\epsilon^{abc}\epsilon^{def}\epsilon^{a'b'c'}\epsilon_{jkl} E^i_c 
E^i_f E^j_{b'}E^k_{c'} \nn \\
& & \qquad \qquad  \big[ E^m_b(E^l_{d,a}E^m_{e, a'}  +  
E^l_{e,a}E^m_{a', d})   + E^m_e E^l_{d,a}(E^m_{b, a'} - E^m_{a',b})  
\big] \label{C^g_3} \quad .
\ee
This means that the ``potential''--term in the Hamiltonian constraint is 
symbolically  of the form
\be
\det (E) R(E) & \propto & \frac{E^2 (\partial E)^2}{\det (E)} + 
\frac{E^3 (\partial \partial E)}{\det (E)} + \frac{E^5 (\partial 
E)^2}{(\det (E))^2} \quad .
\ee
In terms of the momenta $P^a_i$, the cotriad $E_a^i$ and its determinant 
the diffeomorphism constraint can be written as
\be
C_a^g & = & -E_a^i D_b P^b_i = - E_a^i \partial_b P^b_i -  
\epsilon_{ijk}E_a^i \Gamma_b^j P_k^b \nn \\
& = & -  E_a^i \partial_b P^b_i + \frac{1}{2}P^b_i(E_{a,b}^i - 
E_{b,a}^i) + \frac{1}{4 \det (E)}\epsilon^{cdg}\epsilon_{ijk}E_d^j E_g^k 
P^b_i \left[ E_a^l (E_{b,c}^l - E_{c,b}^l) + E_b^l (E_{a,c}^l - 
E_{c,a}^l)  \right] \quad .
\ee
Throughout this article we will work in units where $c=1$ and the 
coordinates of $\Sigma$ have the dimension of length i.e. $[\sigma] = cm$. Thus, 
the metric $Q_{ab}$ is dimensionless as well as the cotriads, $[E_a^i] = 
1$. The extrinsic curvature of $\Sigma$ is measured in units of 
$cm^{-1}$ which directly carries over to our canonical momenta, $[P^a_i] 
= cm^{-1}$. The gravitational coupling constant carries units $[\kappa] 
= cm \cdot kg^{-1}$ and we will leave Planck's constant dimensionful, 
$[\hbar] = kg \cdot cm$.

\subsection{Coupling matter} \label{sec:matter}

Now we want to couple matter to the gravitational sector. In principle 
one can allow all (standard model) matter but for illustrative purposes 
we want to restrict ourselves to the case of a minimally coupled scalar 
field of Klein--Gordon type in this article: The basic variables are 
given by a scalar field $\Phi$ and its canonically conjugate momentum 
$\Pi$, both assumed to be in $C^\infty({\Sigma})$:
\be
\{ \Pi(\sigma), \Phi(\sigma)  \} = \lambda \delta(\sigma, \sigma')
\ee
with the matter coupling constant $\lambda$ which we will set equal to 
one for the remainder of this article.\\
In ADM--variables the Hamiltonian and diffeomorphism constraints in the 
matter sector are given by
\be
C^\phi & = &  \frac{\Pi^2}{2\lambda\sqrt{ \det Q}} + \frac{\sqrt{\det 
Q}}{2 \lambda}(Q^{ab}\partial_a\Phi \partial_b \Phi + 
\frac{m^2}{\hbar^2} \Phi^2)\\
C^\phi_a & = & \Pi \partial_a \Phi
\ee
respectively which can directly be rewritten in terms of our 
gravitational variables $(E_a^i, P^a_i)$ as
\be
C^\phi & = &  \frac{1}{2\lambda \det (E)} \Pi^2 + \frac{1}{2 \lambda 
\det (E)}\epsilon^{acd}\epsilon^{bef} E^k_c E^l_d E^k_e E^l_f \Phi_{,a} 
\Phi_{,b}  + \frac{1}{2 \lambda}\det (E) \frac{m^2}{\hbar^2} \Phi^2 \\
C^\phi_a & = & \Pi \partial_a \Phi \quad . 
\ee
In the above definitions we keep the factor of $\frac{1}{\hbar^2}$ in 
the definition of the massive term to make sure that $\Pi, \Phi$ have 
the right dimensions: It can easily be seen that $[\Phi] = kg^{-1/2} 
\cdot cm^{-1/2}$ and $[\Pi] = kg^{1/2} \cdot cm^{-3/2}$.\\
{The physical Hamiltonian, which} generates evolution in dust--time for 
the whole system gravity plus matter is now given by
\be
H_{\rm phys} = \int\limits_{\Sigma}d\sigma\sqrt{(C^g + C^\phi)^2 - 
Q^{ab} (C^g_a + C^\phi_a)(C^g_b + C^\phi_b)} \quad .
\ee
After the canonical transformation in the gravitational sector the 
matter degrees of freedom only couple to the cotriads $E_a^i$ and not to 
their momenta $P^a_i$. When turning to the quantum theory in section 
\ref{sec:quantum} we will choose a representation where the cotriads are 
quantised as multiplication operators as opposed to the 
LQG-representation where (smeared) densitised triads act as 
(non--commuting) right--invariant vector fields on SU(2). Albeit loosing 
the non--commutative structure (which might turn out to be important in 
the long run) this is one of the main advantages of working in this 
representation because having to deal only with gravitational 
multiplication operators in the analysis of the matter sector makes it 
much easier to analyse the QFT on CS limit in this framework.\\

\subsection{Discretisation} \label{sec:discretization}

Having rewritten general relativity in terms of the cotriad $E_a^j$ and 
its canonical momentum $P^a_j$ we now want to explain how one can 
discretise this theory on a regular cubic lattice in 3 spatial 
dimensions. This will result in a classical dynamical system of an at 
most countable number of degrees of freedom and it is exactly this 
dynamical system which we will compare the quantum theory with that we 
will define in subsequent sections.\\
Let $\gamma$ be a piecewise analytic graph, embedded in ${\Sigma}$ and 
of cubic topology (that is, each vertex is 6--valent). For details 
concerning piecewise analytic graphs of the type usually employed in LQG 
see for instance \cite{LOST}. For our purposes it is 
sufficient\footnote{We will be rather informal here: Of course one 
should not consider paths as we do here but certain equivalence classes 
thereof. But in this article these details play a minor role so we will 
drop them for the sake of clarity. } to think of $\gamma$ as a 
collection $V(\gamma)$ of points $v \in {\Sigma}$, called vertices, 
together with analytic edges $c_I[v]: (0,1) \rightarrow {\Sigma}; \quad  
s \mapsto c_I^a[v](s)$ such that $c_I[v](0) = v$ and $c_I[v](1) = v + 
I$. The requirement of cubic topology means that each vertex $v$ has 
exactly 3 outgoing edges, that is, $I,J,K = 1,2,3$. Hence, each vertex 
$v$ has exactly 6 neighbouring vertices, which we will denote by $v\pm 
(1,0,0), v\pm (0,1,0)$ and $v\pm (0,0,1)$, or simply as $v \pm I$ for 
the neighbouring vertex in direction $I$.\\
\begin{figure}
\centering{
	\psfrag{I}[tr][tr]{$c_1[v]$}
	\psfrag{J}[bl][bl]{$c_2[v]$}
	\psfrag{K}[br][br]{$c_3[v]$}
	\psfrag{v}[Bc][Bc]{$v$}
	\psfrag{v +I}[tr][tr]{$v + (1,0,0)$}
	\psfrag{v-I}[bl][bl]{$v - (1,0,0)$}
	\psfrag{v+J}[Bl][Bl]{$v + (0,1,0)$}
	\psfrag{v-J}[Br][Br]{$v - (0,1,0)$}
	\psfrag{v+K}[bc][bc]{$v + (0,0,1)$}
	\psfrag{v-K}[tc][tc]{$v - (0,0,1)$}
	\includegraphics[scale=0.5]{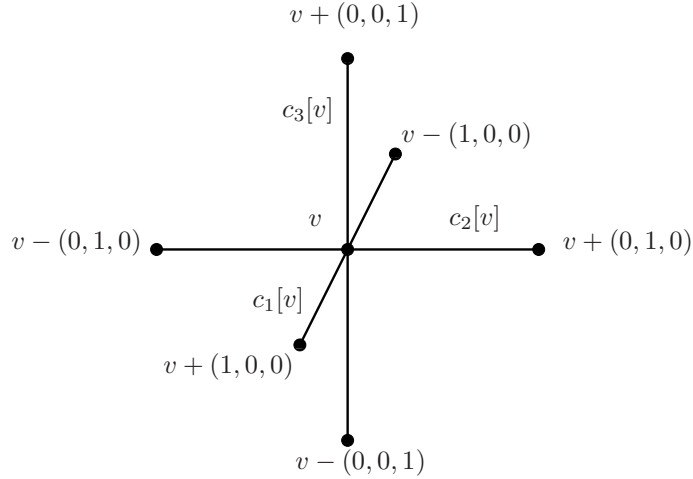}
	\caption{A part of the graph $\gamma$ with vertices $v, v\pm 
(1,0,0), v\pm (0,1,0), v\pm (0,0,1)$ and (outgoing) edges $c_1(v), 
c_2(v), c_3(v)$.}
}
\end{figure}
Furthermore we will need the dual graph $\gamma^*$ whose faces are 
defined through a smooth embedding
\be
S_{c_I[v]}: (0,1) \times (0,1) \rightarrow {\Sigma}; \quad s_1, s_2 
\mapsto S_{c_I[v]}(s_1, s_2) \quad .
\ee 
The embedding is chosen in such a way that each surface $S_{c_I[v]} \in 
\gamma^*$ intersects $\gamma$ only in exactly one point $p_c$, which is 
an interior point of $S_{c_I[v]}$ as well as $c_I[v]$.\\
\begin{figure}
\centering{
	\psfrag{gamma}[Bl][Bl]{$\gamma$}
	\psfrag{gammastar}[Bl][Bl]{$\gamma^*$}
	\includegraphics[scale=0.5]{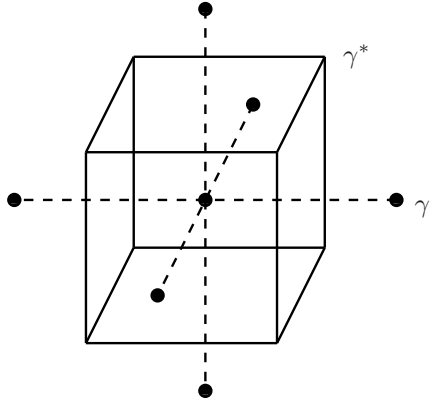}
	\caption{A part of the graph $\gamma$ and the corresponding part 
of its dual graph $\gamma^*$.}
}
\end{figure}
A discretised version of general relativity using the variables defined 
above should not depend on any background metric, since there simply is 
none. In order to define lattice variables that do not show any 
dependence on a metric one proceeds as follows: $E_a^j$ is an 
$su(2)$--valued one--form, so we can naturally (that is without 
reference to any metric) integrate it along one--dimensional 
submanifolds of ${\Sigma}$, namely the edges. $P^a_j$ is an 
$su(2)$--valued vector density, so we can naturally integrate it over 
the faces of the dual graph. Similiar reasoning holds for the matter 
variables: $\Pi$ is a scalar density, thus it can naturally be 
integrated over the volumina of the dual graph, and for $\Phi$ as a 
scalar field of density weight zero the natural thing to do is to just 
take the value of the field at each vertex.
Thus, we define
\be
E_I^i(v) & := & \int\limits_0^1 ds \partial_s c_I^a[v](s) 
E_a^i(c_I[v](s)) \quad ,  \\
P^I_i(v) & := & \int\limits_0^1ds_1\int\limits_0^1ds_2 \frac{\partial 
S_{c_I[v]}^a}{\partial s_1 }\frac{\partial S_{c_I[v]}^b}{\partial s_2} 
\epsilon_{abc} P^c_i(S_{c_I[v]}(s_1,s_2)) \quad , \\
\Phi(v) & := & \Phi(\sigma)|_{\sigma = v} \quad ,\\
\Pi (v) & := & \int\limits_{V(v)} d\sigma \Pi(\sigma)  \quad .
\ee
These discrete variables fulfil the Poisson algebra\footnote{Here we 
fix the matter coupling constant $\lambda =1 $.}
\be 
\{ P_i^I(v), E_J^j(v') \} & = & \kappa \delta_J^I \delta_i^j 
\delta_{vv'} \quad , \nn \\
\{ P_i^I(v), P_j^J(v')  \} = \{ E_I^i(v), E_J^j(v')  \} & = & 0 \quad , 
\nn \\
\{ \Pi(v), \Phi(v')  \} & = & \delta_{vv'} \quad , \nn \\
\{ \Pi(v), \Pi(v') \} = \{ \Phi(v), \Phi(v')  \} & = & 0 \quad 
.\label{discrete_poisson}
\ee
Poisson brackets between matter and geometry variables vanish obviously 
and the $\delta_{vv'}$ appearing in the above formulas is now the 
Kronecker delta. The Poisson brackets lose their distributional form 
exactly because we have chosen to smear the dynamical variables in 
dimensions which add up to $\dimm({\Sigma})$ for canonical pairs.
Due to the smearing procedure the discrete variables carry different 
units compared to the continuum variables, namely $[E_I^i(v)] = cm$, 
$[P^I_i(v)] = cm$, $[\Phi(v)] = kg^{1/2} \cdot cm^{-1/2} $  and 
$[\Pi(v)] = kg^{1/2} \cdot cm^{3/2}$.\\
Now we can define discrete analogues of the constraints from the last 
section by simply replacing
\be
E_a^i(\sigma) & \rightarrow & E_I^i(v) \quad , \nn \\
P_i^a(\sigma) & \rightarrow & P^I_i(v) \quad , \nn \\
\Phi(\sigma) & \rightarrow & \Phi(v) \quad , \nn \\
\Pi(\sigma) & \rightarrow & \Pi(v) \quad ,
\ee
and replacing the integrals $\int\limits_{{\Sigma}}d\sigma$ by Riemann 
sums $\sum\limits_{V(\alpha)}$.
Of course there is an ambiguity in defining these  discrete analogues
just because there is no unique way of replacing the derivatives of 
$E_a^i(\sigma)$: One could replace $\partial_a E_b^i(\sigma) \rightarrow 
\nabla_I^+ E_J^i(v)$ or  $\partial_a E_b^i(\sigma) \rightarrow 
\nabla_I^- E_J^i(v)$ where $\nabla_I^+, \nabla_I^-$ are the lattice 
forward and backward derivatives into direction $I$ respectively defined 
through $\nabla_I^+f(v) := f(v+I)-f(v)$ and $\nabla_I^- f(v) := f(v) - 
f(v - I)$ for functions $f: V(\gamma) \rightarrow \TC$ (in our case 
either $\TC = \RR$ or $\TC = su(2)$ but one could have different 
topological spaces $\TC$). Both definitions (and {others involving not 
only next -- neighbouring terms}) are equivalent when taking the 
continuum limit.\\
\\
Remark:\\
The algebra of the naively discretised Hamiltonian densities or any 
other quantities involving spatial derivatives that are replaced by 
difference operators will differ 
from their continuum limits even in the classical theory. This poses no 
consistency problem because 
these are no constraints anymore but true Hamiltonian densities, there 
are no anomalies. However, the densities will no longer be constants of 
the motion and one may want to improve the situation. If we would have 
constraints then one could fix the corresponding Lagrange multipliers by 
the method proposed in 
\cite{gambini_pullin1, gambini_pullin2, 
uniform_discretizations1, uniform_discretizations2}. However, we are not 
in this situation and thus one would need to rather change the 
naive discretisation into a more subtle one as suggested in 
\cite{AQGIV}, possibly along the lines of \cite{Hasenfratz}.
Notice that we still do have a constraint, namely the Gauss constraint,
but in our variables, in contrast to the connection variables, this does 
not involve spatial derivatives and thus there are no anomalies.

\section{Quantum theory} \label{sec:quantum}

\label{sec:algebraic}

In this section we want to define a quantum  theory of   {General 
Relativity} in the formulation developed in section \ref{sec:classical}. 
To be more specific, we will start with the algebra 
(\ref{discrete_poisson}) and explain how one can represent it on an 
appropriately chosen Hilbert space. 
 {
First we will define a quantum theory on a {\it purely algebraic} level, 
that is, we will start with an elementary algebra of observables $\AK$ 
labeled by an algebraic graph $\alpha$ and then construct a 
representation of this algebra on some Hilbert space $\HC$ defined on 
that graph. Thus, dynamics will be defined on an abstract graph, which 
does not know anything about embedding, differential or even topological 
structures of a manifold, since there is no manifold at this stage. 
In the framework of LQG many things do actually not depend on the 
embedding and this was one of the motivations in \cite{AQGI} to 
formulate an LQG inspired quantum theory purely on the algebraic level.}
However, if we want to interpret this theory as a quantum theory of  
{General Relativity} we need to make sure that there exists a large 
enough class of semiclassical states (and these states will carry 
information about the embedding of the graph $\alpha$ into the manifold 
${\Sigma}$ to be approximated) and it can indeed be shown that such 
semiclassical states exist (see section \ref{sec:classical_limit}). If 
the quantum system is in such a state the expectation values of 
geometrical observables will be sharply peaked on classical values and 
thus, we are getting classical discretised gravity as a classical limit 
of our 
algebraic theory.
\vspace{.2cm}\\
Now we want to define the quantum theory living on an algebraic graph 
and we will do this in a similar way as in \cite{AQGI}. First we need to 
define what we mean by an algebraic graph as opposed to the embedded, 
piecewise analytic graph $\gamma$ from section \ref{sec:discretization}: 
Let $\alpha$ be an oriented algebraic graph with N vertices, i.e. a 
countable set $V(\alpha)$ of $N$ points (``vertices'') together with an 
$N \times N$ matrix $\alpha$ whose entries $\alpha_{vv'}, v,v' = 
1,\dots, N$ take non--negative integer values. In general $\alpha_{vv'}$ 
is not a symmetric matrix for an oriented graph and the total number of 
``edges'' connecting vertex $v$ and vertex $v'$ is given by $c_{vv'} := 
\alpha_{vv'} + \alpha_{v'v}$. To be more intuitive, one can think of 
$\alpha$ as all the information that remains if one starts with an 
embedded, piecewise analytic graph $\gamma$, consisting of $N$ vertices 
and $\alpha_{vv'}$ embedded edges between vertex $v$ and $v'$  and then 
drops everything that $\gamma$ knows about metric, differentiable or 
even topological structures of ${\Sigma}$. $\alpha$ neither contains any 
information about the braiding of its edges nor about smoothness or 
$n$--times differentiability usually associated to edges which are 
embedded in a manifold.\\
For the remainder of this article we will be interested in the case $N = 
\aleph$ and graphs of cubic topology, i.e. each vertex $v \in V(\alpha)$ 
is 6 valent and $c_{vv'} \in \{0,1 \}\, \forall v, v'$. Thus, each 
vertex $v$ has three outgoing edges, which we will label by 
$I,J,K=1,2,3$ for bookmarking purposes and three ingoing edges which 
will be counted as outgoing edges for the respective neighbouring vertex 
in direction $-I,-J,-K$.\\

\subsection{The algebra}
Now we define an abstract $^\ast-$algebra $\AK$ 
whose generators are labeled by vertices of the algebraic graph $\alpha$
and which implements the $^\ast-$relations and canonical commutation 
relations that arise from the classical reality conditions (all 
discretised fields are real valued) and the 
Poisson brackets (\ref{discrete_poisson}):
\subsubsection{Gravitational sector}
Let $\alpha$ be the cubic algebraic graph with $N=\aleph$ as described 
above. With each vertex $v$ of $\alpha$ associate a triple $E_I^i(v)$ of 
$su(2)$--valued quantities where $I,J,K,\dots$ label the 3 outgoing 
edges at each vertex and $i,j,k,\dots$ are $su(2)$--indices. Furthermore 
associate a triple of $su(2)$--valued quantities $P^I_i(v)$ with each 
vertex where again capital letters denote outgoing edges and small 
letters are $su(2)$--indices. Let these basic variables be subject to 
the following algebraic relations:
\be
\big[ E_I^i(v), E_J^j(v') \big]  & = & 0 \nn \\
\big[ P^I_i(v), P^J_j(v') \big] & = & 0 \nn \\
\big[ P^I_i(v), E_J^j(v') \big] & = & \kappa 
\delta_{IJ}\delta_{ij}\delta_{vv'} \quad.
\ee
Furthermore we can define an involution on the algebra and  will demand 
the (trivial) $\,^*$--relations
\be
(E_I^i(v))^* = E_I^i(v), \quad (P^I_i(v))^* = P^I_i(v) \quad .
\ee
Thus  $E_I^i(v)$ and $P^I_i(v)$ form an abstract $\,^*$--algebra with 
the above relations which we want to denote as $\mathfrak{A}^g$. 

\subsubsection{Matter sector}
The same thing can be done for the matter degrees of freedom: With each 
vertex $v$ of $\alpha$ associate a pair of real valued quantities 
$\Phi(v)$ and $\Pi(v)$. Let these variables be subject to the algebraic 
relations
\be
\big[ \Phi(v), \Phi(v')  \big] & = & 0 \nn \\
\big[ \Pi(v), \Pi(v')  \big] & = & 0 \nn \\
\big[ \Pi(v), \Phi(v')  \big] & = & \delta_{vv'} \quad ,
\ee
and again define a (trivial) involution $\,^*$ through
\be
(\Phi(v))^* = \Phi(v), \quad (\Pi(v))^* = \Pi(v) \quad .
\ee
Again $\Phi(v)$ and $\Pi(v)$ together with the above relations form an 
abstract $\,^*$--algebra which we want to denote by 
$\mathfrak{A}^\phi$.\\
So the whole theory is described by an abstract $\,^*$--algebra 
$\mathfrak{A} := \mathfrak{A}^g \otimes \mathfrak{A}^\phi$.

\subsection{The Hilbert space}
\label{Hspace}

$\mathfrak{A}$ can be represented as an algebra of linear operators on 
an ITP Hilbert space $\mathcal{H}_\otimes := \otimes_v \mathcal{H}_v$
where $\mathcal{H}_v=\mathcal{H}^g_v\otimes \mathcal{H}^m_v$. 
where the individual Hilbert spaces on each vertex, $\mathcal{H}^g_v$ 
and $\mathcal{H}^\phi_v$, are chosen as explained below. 
Note that this ITP is different from
\be
\mathcal{H}^g_\otimes \otimes 
\mathcal{H}^\phi_\otimes : = (\otimes_v \mathcal{H}^g_v) \otimes 
(\otimes_v \mathcal{H}^\phi_v) \quad , 
\ee 
because of the non associativity of the infinite tensor product. 
The bracketing chosen by us is preferred because it allows to construct 
gauge invariant quantities. More information about the ITP construction 
can be found in 
appendix \ref{ITP}.

\subsubsection{Gravitational sector}
Let $\mathcal{H}^g_v := L_2(\mathbb{R}^9, d\mu)$ be a Hilbert space 
associated to each vertex $v$ with Lebesgue measure $\mu$ on $\RR^9$ and 
consider the ITP Hilbert space $\mathcal{H}^g_\otimes$ . On 
$\mathcal{H}^g_\otimes$ the algebra $\mathfrak{A}^g$ can be represented 
by choosing  a representation $\rho: \mathfrak{A}^g \rightarrow 
\LC(\HC^g_\otimes)$ in which $E_I^i(v)$ acts as a multiplication 
operator and $P^I_i(v)$ as a derivative operator,
\be
\hat{E}_I^i(v)\cdot \psi(E) := \big[ \rho(E_I^i(v)) \psi \big](E) & := & 
E_I^i(v) \cdot \psi(E) \\
\hat{P}^I_i(v) \cdot \psi(E) := \big[ \rho(P^I_i(v)) \psi \big](E) & := 
& i\ell_P^2 \frac{\partial}{\partial E_I^i(v)} \psi(E) \quad ,
\ee
for all $\psi(E) \in \HC^g_\otimes$ and $\ell_P = \sqrt{\kappa \hbar} $ 
is the Planck length\footnote{We need to include a factor of $\ell_P^2$ 
into the definition of the operator $P^I_i(v)$ in order to get the same 
dimensions as its lattice--analogue.}. Note that the derivative in the 
second line is indeed just the partial derivative and not a functional 
derivative. As operators on $\HC_\otimes^g$ the cotriad $E_I^i(v)$ and 
its conjugate momentum  $P^I_i(v)$ fulfil the commutator relations
\be
\big[ \hat{P}^I_i(v), \hat{E}_J^j(v') \big] & = & i \ell_P^2 
\delta_{IJ}\delta_{ij}\delta_{vv'} \quad.
\ee
In what follows we will mostly omit the hat above the operators and use 
the same symbols  $E_I^i(v), P^I_i(v)$ for elements of the abstract 
$\,^*$--algebra $\AK^g$ and their representatives as linear operators on 
the Hilbert space $\HC^g_\otimes$ but it should be clear from the  
context which one we mean.

\subsubsection{Matter sector}
Again, a natural representation of $\mathfrak{A}^\phi$ is given by the 
ITP representation on $\mathcal{H}^\phi := \otimes_v \mathcal{H}^\phi_v$ 
where each vertex labels a Hilbert space $\mathcal{H}_v^\phi := 
L_2(\mathbb{R}, d\nu)$ with $\nu$ being the Lebesgue measure on 
$\mathbb{R}$. We choose a representation such that
\be
\hat{\Phi}(v)\cdot \psi(\Phi) & := & \Phi(v) \psi(v) \quad , \\
\hat{\Pi}(v) \cdot \psi(\Phi) & := & i \hbar \partial_{\Phi(v)} 
\psi(\Phi) \quad ,
\ee
for $\psi \in \HC^\phi_\otimes$. As linear operators on 
$\HC_\otimes^\phi$ these fulfil the commutator relations
\be
[\hat{\Pi}(v), \hat{\Phi}(v)  ] = i \hbar \delta_{vv'} \quad .
\ee

\subsection{The Hamiltonian Density}
In section \ref{sec:approx} we saw that, given that $\frac{ Q^{ab} C_a 
C_b}{C^2}$ is small, one can approximate the physical Hamiltonian by 
$H_{\rm phys} \approx \int\limits_{\Sigma}d\sigma \left(C^g(\sigma) + 
C^\phi(\sigma)\right)$ plus higher order terms. So in order to define 
physical dynamics in the quantum sector, the first step is to have a 
well defined quantum operator $\hat{C}$ associated to the non 
dust contributions of the classical 
Hamiltonian constraint  $\int\limits_{\Sigma}d\sigma \left(C^g(\sigma) + 
C^\phi(\sigma)\right)$. For higher order corrections we will need the 
diffeomorphism constraint as well, so for completeness we will also give 
a 
quantum operator $\hat{C}_a(v)$ associated to the non dust 
contributions to the classical 
diffeomorphism constraint density.\\
The crucial step in defining these operators is to be able to define 
operator analogues for functions involving inverse powers of $\det(E)$.  
In this section we will just formally define these operators, in the 
next section we will show in detail that they are in fact symmetric 
operators on a dense subspace of $\mathcal{H}_\otimes$.\\
All the operators will be of the form $O = O^g \otimes O^\phi$, where 
the first part, $O^g$, acts only on $\mathcal{H}^g_\otimes$ and the 
second part, $O^\phi$, acts only on $\mathcal{H}^\phi_\otimes$.

\subsubsection{Gravitational sector}
As the operator analogue for the (integrated) gravitational part of the 
Hamiltonian constraint we define\footnote{For the most part of what 
follows we will again drop the hats over the constraint operators 
when there is no ambiguity in mistaking them for the classical 
functions.}
\be
\hat{C}^g & := & (\hat{C}^g_{\rm kin} - \hat{C}^g_{\rm pot}) \otimes 
id_\phi \label{C_g_q}\\
\hat{C}^g_{\rm pot} & := & \hat{C}^g_{\rm pot_1} + \hat{C}^g_{\rm pot_2} 
+ \hat{C}^g_{\rm pot_3}\\
\hat{C}^g_{\rm kin} & := & \kappa \sum\limits_{v} \sum\limits_{I,J} 
\left[ i \ell_P^2 \partial_{E_I^i(v)} \right]G_{IJ}^{ij}\left[ i 
\ell_P^2 \partial_{E_J^j(v)} \right] \quad , \\ 
\hat{C}^g_{\rm pot_1} & := & \frac{1}{4\kappa \det E(v) } 
\sum\limits_{I,J,K,L,M,N} \epsilon^{IJK}\epsilon^{LMN} E_I^k(v) E_L^l(v) 
\big( \nabla_K^+ E_J^k(v) \big) \big( \nabla_N^+ E_M^l(v) \big) \quad , 
\\
\hat{C}^g_{\rm pot_2} & := & -\frac{1}{4\kappa  \det E(v) 
}\sum\limits_{I,J,K,L,M,N} \epsilon^{IJK}\epsilon^{LMN} \nonumber \\
& & \quad  \left[ E_K^k(v) \big( \nabla_J^+ E_L^k(v) - \nabla_L^+ 
E_J^k(v) \big) + E_L^k(v)\big( \nabla_J^+ E_K^k(v) \big) \right]\times  
\nn \\
& & \quad \times \left[ E_N^k(v) \big( \nabla_M^+ E_I^k(v) - \nabla_I^+ 
E_M^k(v) \big)  + E_I^l(v) \big( \nabla_M^+ E_N^l(v) \big) \right] \quad 
, \\
\hat{C}^g_{\rm pot_3} & := & -\frac{1}{ \kappa  \det E(v) } 
\sum\limits_{I,J,K,L,M,N} \epsilon^{IJK}\epsilon^{LMN} E_K^k(v)E_N^k(v) 
\times \nonumber \\
& & \quad  \times \left[ E_M^l(v) \nabla_I^- \nabla_L^+ E_J^l(v) + \big( 
\nabla_I^+ E_J^l(v) \big) \big( \nabla_L^+ E_M^l(v) \big) + E_J^l(v) 
\big( \nabla_I^- \nabla_L^+ E_M^l(v) \big)  \right] \nonumber \\
& & + \frac{1}{2 \kappa \big( \det E(v) \big)^2} 
\sum\limits_{I,J,K,L,M,N,O,P,Q} 
\epsilon^{IJK}\epsilon^{LMN}\epsilon^{OPQ} \epsilon_{jkl} 
E_K^i(v)E_N^i(v)E_P^j(v)E_Q^k(v) \times \nonumber \\
& & \quad \times \left[ E_J^m(v)\big[ \big( \nabla_I^+ E_L^l(v) \big) 
\big( \nabla_O^+ E_M^m(v) \big) + \big( \nabla_I^+ E_M^l(v) \big) \big( 
\nabla_L^+ E_O^m(v)\big) \big] \right. \nonumber \\
& & \quad \quad + \left. E_M^m(v) \big( \nabla_I^+ E_L^l(v) \big) \big[  
\big( \nabla_O^+ E_J^m(v) \big)  -  \big( \nabla_J^+ E_O^m(v) \big)    
\big] \right] \quad .
\ee
$id_\phi$ is the identity operator on $\HC_\otimes^\phi$. The 
multiplication operator in the kinetic term is given by
\be
G_{IJ}^{ij}[E(v)] := \frac{1}{ 4 \det E(v)} (E^k_I(v) E^k_J(v) 
\delta^{ij}- \frac{1}{2}E_I^i(v) E_J^j(v))
\ee
and by $\nabla_I^+$ and $\nabla_I^-$ we denote the lattice forward and 
backward derivatives into direction $I$ respectively defined through 
$\nabla_I^+ f(v) := f(v+I) - f(v)$ and $\nabla_I^- f(v) := f(v) - f(v - 
I)$ for functions $f: V(\alpha) \rightarrow su(2)$.\\
The operator $\hat{C}^g_I(v) =   \hat{\hat{C}}^g_I(v) \otimes id_\phi$, 
associated to the classical gravitational diffeomorphism constraint 
density, can be defined as
\be \label{C_I_g_q}
\hat{\hat{C}}^g_I(v) & := & -\frac{1}{2}\sum\limits_{J,j} E_I^j 
(\nabla_J^+ P_j^J(v)) + (\nabla_J^+ P_j^J(v)) E_I^j(v) \\
& & + \frac{1}{4}\sum\limits_{J,j} P^J_j(v) \left[ (\nabla_J^+ E_I^j(v)) 
- (\nabla_I^+ E_J^j(v))  \right] +  \left[ (\nabla_J^+ E_I^j(v)) - 
(\nabla_I^+ E_J^j(v))  \right]P^J_j(v) \nn \\
& & + \frac{1}{8}\sum\limits_{\stackrel{JKLM}{ijkl}}\epsilon^{KLM} 
\epsilon_{ijk} \Big[ \nn \\
& & \quad  P_i^J(v)\left[ \frac{1}{\det (E)} E_L^j(v)E_M^k(v)   \Big[ 
E_I^l(v) \Big( (\nabla_K^+ E_J^l(v)) - (\nabla_J^+ E_K^l(v))   \Big) + 
E_J^l(v) \Big( (\nabla_K^+ E_I^l(v)) - (\nabla_I^+ E_K^l(v))  \Big)  
\Big]  \right] \nn \\
& & + \left. \left[ \frac{1}{\det (E)} E_L^j(v)E_M^k(v)   \Big[ E_I^l(v) 
\Big( (\nabla_K^+ E_J^l(v)) - (\nabla_J^+ E_K^l(v))   \Big) + E_J^l(v) 
\Big( (\nabla_K^+ E_I^l(v)) - (\nabla_I^+ E_K^l(v))  \Big)  \Big]  
\right] P_i^J(v) \right] \quad . \nn
\ee
Here we used the notation $\nabla_I^+ P^J_j(v) := P^J_j(v+I) - P^J_j(v) 
= i\ell_P^2 \partial_{E_J^j(v+I)} - i \ell_P^2 \partial_{E_J^j(v)}$. 
This operator, besides the terms proportional to $\nabla_I^+ E_J^j(v)$, 
causes $\hat{C}^g_I(v)$ to act not only on vertex $v$ but also on its 
neighbouring ones.\\
Finally the gravitational $SU(2)$--Gauss constraint operator can be 
defined as
\be
\hat{G}[\Lambda] & = & \id_\phi \otimes \frac{1}{2}\sum\limits_{v \in 
V(\alpha)} \epsilon_{ijk} \left[  P^I_j(v) E_I^k(v) \Lambda_i(v) +   
E_I^k(v) \Lambda_i(v) P^I_j(v) \right]
\ee
for $\Lambda: V(\alpha) \rightarrow SU(2); \quad v  \mapsto 
\Lambda_i(v)$.

\subsubsection{Matter sector}
As operator analogues for the (integrated) matter Hamiltonian constraint 
and the matter diffeomorphism constraint density we define:
\be
\hat{C}^\phi & := & \frac{1}{2}\sum\limits_{v}\Big[ \frac{1}{\det E(v)} 
\otimes \Pi^2(v) \nonumber \\
& & \quad \quad  \quad + \sum\limits_{I,J,K,L,M,N}\frac{1}{2 \det 
E(v)}\epsilon^{IKL}\epsilon^{JMN} 
E_K^k(v)E_L^l(v)E_M^k(v)E_N^l(v)\otimes (\nabla_I^+ \Phi(v)) 
(\nabla_J^+\Phi(v))\nonumber \\
& & \quad \quad \quad + \det E(v) \otimes \frac{m^2}{\hbar^2} \Phi^2(v) 
\quad \Big]  \label{C_phi_q}\\
\hat{C}^\phi_I(v) & := & id_g \otimes \frac{i \hbar}{2}\left[  
(\nabla_I^+\Phi(v)) \partial_{\Phi(v)} + 
\partial_{\Phi(v)}(\nabla_I^+\Phi(v)) \right] \label{C_I_phi_q} \quad .
\ee
One can rewrite (\ref{C_phi_q}) in a slightly different way which will 
become more convenient for the analysis of the matter sector which we 
want to perform later on
\be
\hat{C}^\phi & = & \frac{1}{2} \sum\limits_{v \in V(\alpha)} \Pi(v) 
\frac{1}{\det E(v)} \Pi(v) + \Phi(v) \det E(v) (-\Delta + 
\frac{m^2}{\hbar^2})\Phi(v) \quad ,
\ee
with the lattice Laplace--Beltrami operator $\Delta$ defined as
\be
\Delta & = & \frac{1}{\det E(v)} \sum\limits_{I,J,K,L,M,N}\nabla_I^- 
\left(\frac{1}{2 \det E(v)} \epsilon^{IKL} \epsilon^{JMN} E_K^k(v) 
E_L^l(v) E_M^k(v) E_N^l(v) \nabla_J^+\right) \quad .
\ee
Here we used that $\nabla_I^+ = (\nabla_I^-)^\dagger$ if interpreted as 
an operator on $l_2(V(\alpha))$, the Hilbert space of square--summable 
functions on the set of vertices $V(\alpha)$. This Hilbert space, 
$l_2(V(\alpha))$, should not be mistaken for the ITP Hilbert space which 
we based our quantum theory on. But it will play a role as the ``one 
particle'' Hilbert space in the construction of an appropriate Fock 
space.

\subsection{Domains of Definition}\label{sec:operators}

In order to show that the operators written down in the last section are 
actually well defined symmetric operators on a dense subspace of 
$\mathcal{H}_\otimes$ we proceed as follows: As can easily be seen the 
parts acting on $H^\phi_\otimes$ will cause no problems. For each vertex 
$v$ we can define the matter part of (\ref{C_phi_q}) and 
(\ref{C_I_phi_q}) on the Schwarz space $\mathcal{S}(\mathbb{R})$ of 
smooth functions of rapid decrease which 
is a dense subspace 
of $\mathcal{H}^\phi_v = L_2(\mathbb{R}, d\nu)$.\\
In order to show that the operators (\ref{C_g_q}), (\ref{C_I_g_q}) as 
well as the gravitational part of (\ref{C_phi_q}) and (\ref{C_I_phi_q})
is well defined it is sufficient to show that $A^n(v) := \frac{1}{(\det 
(E)(v))^n}$ for $n = 1,2$ can be defined on a dense subspace of 
$\mathcal{H}^g_v = L_2(\mathbb{R}^9, d\mu)$ and that the entire 
operators are symmetric on that subspace.\\
Let us analyse the determinant of the cotriad in more detail:
The operator $E_I^i(v)$ (associated to the cotriad) is densely defined 
on $\mathcal{S}(\mathbb{R}^9)$, so we need to show that this holds also 
for powers of the inverse of its determinant . If we view the 
determinant $\det: \mathbb{R}^9 \rightarrow \mathbb{R}; \quad E_I^i(v) 
\mapsto \det (E)(v)$ as a real--valued function on $\mathbb{R}^9$ then 
the singularity structure of $\frac{1}{\det (E)(v)}$ can most easily be 
analysed if we perform a singular value decomposition as follows:
\be
E & =: & L D R \quad ,
\ee
where $L,D \in SO(3)$ and D is a diagonal $3\times 3$--matrix with real 
eigenvalues\footnote{We could have absorbed the signs of the 
eigenvalues into $L,R$ by having them take values in $O(3)$. But it is 
more convenient to have them take values on the whole real axis. 
Notice that we do not insist on definite sign of $\det(E)$ on the whole 
phase space.}.
Explicitly these matrices are given by
\be
E & := &
\begin{pmatrix}
E^1_1 & E_1^2 & E_1^3 \\
E_2^1 & E_2^2 & E_2^3 \\
E_3^1 & E_3^2 & E_3^3
\end{pmatrix}
\\
 & \; & \nonumber
\\
L & := &
\left( \begin {array}{ccc} \cos \left( \beta \right) \cos \left( \gamma 
\right) &-\cos \left( \beta \right) \sin \left( \gamma \right) &\sin 
\left( \beta \right) \\\noalign{\medskip}\sin \left( \alpha \right) \sin 
\left( \beta \right) \cos \left( \gamma \right) +\cos \left( \alpha 
\right) \sin \left( \gamma \right) &-\sin \left( \alpha \right) \sin 
\left( \beta \right) \sin \left( \gamma \right) +\cos \left( \alpha 
\right) \cos \left( \gamma \right) &-\sin \left( \alpha \right) \cos 
\left( \beta \right) \\\noalign{\medskip}-\cos \left( \alpha \right) 
\sin \left( \beta \right) \cos \left( \gamma \right) +\sin \left( \alpha 
\right) \sin \left( \gamma \right) &\cos \left( \alpha \right) \sin 
\left( \beta \right) \sin \left( \gamma \right) +\sin \left( \alpha 
\right) \cos \left( \gamma \right) &\cos \left( \alpha \right) \cos 
\left( \beta \right) \end {array} \right) \nn
\\
& \; & \nonumber
\\
D & := &
\begin{pmatrix}
\lambda_1 & 0 & 0 \\
0 & \lambda_2 & 0 \\
0 & 0 & \lambda_3
\end{pmatrix} \nn
\\
& \; & \nonumber
\\
R & := &
\left( \begin {array}{ccc} \cos \left( \theta \right) \cos \left( 
\varphi  \right) &-\cos \left( \theta \right) \sin \left( \varphi  
\right) &\sin \left( \theta \right) \\\noalign{\medskip}\sin \left( 
\delta \right) \sin \left( \theta \right) \cos \left( \varphi  \right) 
+\cos \left( \delta \right) \sin \left( \varphi  \right) &-\sin \left( 
\delta \right) \sin \left( \theta \right) \sin \left( \varphi  \right) 
+\cos \left( \delta \right) \cos \left( \varphi  \right) &-\sin \left( 
\delta \right) \cos \left( \theta \right) \\\noalign{\medskip}-\cos 
\left( \delta \right) \sin \left( \theta \right) \cos \left( \varphi  
\right) +\sin \left( \delta \right) \sin \left( \varphi  \right) &\cos 
\left( \delta \right) \sin \left( \theta \right) \sin \left( \varphi  
\right) +\sin \left( \delta \right) \cos \left( \varphi  \right) &\cos 
\left( \delta \right) \cos \left( \theta \right) \end {array} \right) 
\nn
\ee
That means we perform a coordinate transformation\\
$\varphi: \mathbb{R}^9 \rightarrow \mathbb{R}^9; \quad (E_1^1, E_1^2, 
E_1^3, E_2^1, E_2^2, E_2^3, E_3^1, E_3^2, E_3^3) \mapsto (\lambda_1, 
\lambda_2, \lambda_3, \alpha, \beta, \gamma, \delta, \theta, \phi)$, 
where the range of the new coordinates is  $\alpha, \gamma, \delta, \phi 
\in (0, 2 \pi],\quad  \beta, \theta \in (0, \pi]$ and $\lambda_1, 
\lambda_2, \lambda_3 \in (-\infty, +\infty)$ and the Jacobian of this 
coordinate transformation can be computed to be
\be
\det(\Phi):=
\det(\partial \varphi/\partial(\lambda,\alpha,..,\phi))  = 
\cos(\beta)\cos(\theta)(-\lambda_3^4\lambda_2^2+\lambda_1^4\lambda_2^2
 -\lambda_1^4\lambda_3^2+\lambda_3^2\lambda_2^4+\lambda_3^4\lambda_1^2
-\lambda_1^2\lambda_2^4) 
\quad .
\ee
$L, R \in SO(3)$, thus the determinant in this coordinate system is 
simply given by
\be
\det (E) = \lambda_1 \lambda_2 \lambda_3 \quad .
\ee
So if we compute $||A^n(v) \psi ||_{L_2}$ for $\psi \in 
\mathcal{S}(\mathbb{R}^9)$ we realise
\be
||A^n(v) \psi ||^2_{L_2} := \int d\Omega\;| \cos(\beta) \cos(\theta)|\; 
d\lambda_1 d\lambda_2 d\lambda_3 \frac{
\lambda_1^4(\lambda^2_2-\lambda_3^2) + \lambda_2^4(\lambda^2_3 - 
\lambda^2_1)+\lambda^4_3(\lambda_1^2 - \lambda_2^2)}{\lambda_1^{2n} 
\lambda_2^{2n} \lambda_3^{2n} } |\Psi(\Omega, \lambda_i)|^2
\ee
where we used $\Omega$ as an abbreviation for all angles and $d\Omega := 
d\alpha d\beta d\gamma d\delta d\theta d\phi$. The denominator in the 
last integral is zero if either $\lambda_1, \lambda_2$ or $\lambda_3$ is 
zero whereas the numerator has its roots at $\lambda_1 = \pm \lambda_2, 
\lambda_2 = \pm \lambda_3$ or $\lambda_1 = \pm \lambda_3$. This means 
that the argument of the integral (including the factor coming from the 
measure) diverges for $\lambda_i \rightarrow 0$ and thus  $A^n(v) := 
\frac{1}{(\det (E)(v))^n}$ does not have finite norm on all 
$\psi \in \mathcal{S}(\mathbb{R}^9)$.
\vspace{.2cm}\\
To cure this problem we need to change the domain of the operator 
$A^n(v)$, and the heuristic idea is to choose as its domain functions 
$\psi \in \mathcal{S}(\mathbb{R}^9)$ with the additional condition that 
$\psi$ vanishes faster than any power of $\lambda_i$ in a neighbourhood 
of the coordinate hypersurfaces defined through $\lambda_i = 0$.\\
In order to do this we will modify the function $\psi$ in a way that 
they have the required properties and still approximate the original 
functions well. For this purpose we adopt a standard regularisation 
procedure. Let us introduce a characteristic function ${\chi}_V$ with 
$V$ being the interval $[-2\epsilon,2\epsilon]$. ${\chi}_V$ is equal to 
one if $x$ lies in $V$ and zero otherwise and is not smooth. 
 Now consider the family of functions
\be
\label{rhoeps}
\rho_{\epsilon}(x)&:=&
\left\{
\begin{array}{ll}C_{\epsilon}
 \exp\left(-\frac{\epsilon^2}{\epsilon^2 - x^2}\right) & {\rm for } 
\quad |x| \leq  \epsilon \\
0 &  {\rm for } \quad |x| \geq \epsilon
\end{array}
\right. \quad .
\ee
The constant $C_{\epsilon}$ is determined by the condition 
$\int_{\RR}dx\rho_{\epsilon} =1$ and is explicitly given by 
$C^{-1}_{\epsilon}=\epsilon\int\limits_{|y|\leq1}\exp(-1/(1-y^2))dy$.
This function is smooth, has compact support since supp 
$\rho_{\epsilon}\subset [-\epsilon,\epsilon]$ and on all of $\RR$ we 
have $\rho_{\epsilon}\geq 0$. We use $\rho_{\epsilon}$
 to regularise the characteristic function ${\chi}_V$ and obtain a 
family of smooth versions of the characteristic function which is equal 
to one if $x$ lies within the interval $[-\epsilon,\epsilon]$ and 
vanishes for all $|x| > 3\epsilon$.
 Let us denote the regularised function as ${\chi}_{\epsilon}$ then
 \be
 \label{chieps}
 {\chi}_{\epsilon}(x) &:=& \int\limits_{\RR} dy {\chi}_{V}(y) 
\rho_{\epsilon}(x-y)= \int\limits_{-2\epsilon}^{2\epsilon} dy  
\rho_{\epsilon}(x-y) 
 \ee
With this definition of $\chi_{\epsilon}$ we can easily show that it 
vanishes if $|x|>3\epsilon$. Since the aim is to set the functions at 
the singularities to zero in a smooth manner we define the function
\be
\label{seps}
s_{\epsilon}(x) &:=&1 -  {\chi}_{\epsilon}(x) =
\left\{
\begin{array}{ll}0  & {\rm for } \quad |x| \leq  \epsilon \\
1 - \int\limits_{-2\epsilon}^{+2\epsilon}dy \rho_{\epsilon}(x-y)  & {\rm 
for } \quad \epsilon \leq |x| \leq  3\epsilon
\\
1 &  {\rm for } \quad |x| \geq 3\epsilon
\end{array}
\right. \quad .
\ee
Notice that $0\le \chi_\epsilon,\;s_\epsilon\le 1$. 
This function is smooth by construction and $s_\epsilon(0) = 0$. Thus, 
we can use this function to modify our original functions $\psi \in 
\mathcal{S}(\mathbb{R}^9)$ such that they vanish at $\lambda_i = 0$ in a 
smooth 
way. Define
\be
\psi_\epsilon: \RR^9 \rightarrow \RR; \quad \Omega, \lambda_i \mapsto 
\psi_\epsilon(\Omega, \lambda_i) = s_\epsilon(\lambda_1) 
s_\epsilon(\lambda_2) s_\epsilon(\lambda_2) \psi(\Omega, \lambda_i)
\ee 
for any $\psi \in \mathcal{S}(\RR^9)$. The product of two smooth 
functions is still smooth, hence the set  $\FC := \{ \psi_\epsilon \}$ 
is a subset of $\mathcal{S}(\RR^9)$ and one can easily see that $\FC$ is 
dense in $\mathcal{S}(\RR^9)$ and therefore also in $L_2(\RR^9)$ in the 
$L_2$--norm:
\be
\label{Psiest}
|| \psi - {\psi}_\epsilon ||_{L_2} & = & \left[\int d\Omega 
\int\limits_{-\infty}^{+\infty} d\lambda_1 
\int\limits_{-\infty}^{+\infty} d\lambda_2 
\int\limits_{-\infty}^{+\infty} d\lambda_3 \det \Phi |  \psi(\Omega, 
\lambda_i) - \psi_\epsilon(\Omega, \lambda_i)|^2 \right]^{\frac{1}{2}} 
\nonumber \\
& = & \left[\int d\Omega \int\limits_{-3\epsilon}^{+3\epsilon} 
d\lambda_1 \int\limits_{-3\epsilon}^{+3\epsilon} d\lambda_2 
\int\limits_{-3\epsilon}^{+3\epsilon} d\lambda_3 \det \Phi |\psi(\Omega, 
\lambda_i)|^2 |1 - 
s_\epsilon(\lambda_1)s_\epsilon(\lambda_2)s_\epsilon(\lambda_3)|^2 
\right]^{\frac{1}{2}} \nonumber \\
& \leq & \left[\int d\Omega \int\limits_{-3\epsilon}^{+3\epsilon} 
d\lambda_1 \int\limits_{-3\epsilon}^{+3\epsilon} d\lambda_2 
\int\limits_{-3\epsilon}^{+3\epsilon} d\lambda_3 \det \Phi |\psi(\Omega, 
\lambda_i)|^2 \right]^{\frac{1}{2}} \nonumber \\
& \leq & \IC_{\rm max} \sqrt{6^3\epsilon^3} \left(\int d\Omega 
|\cos(\beta) \cos(\theta)| \right)^{\frac{1}{2}} \propto 
\epsilon^{\frac{3}{2}} \quad ,
\ee
where $\IC_{\rm max}$ is the supremum of $\left[ |\frac{\det 
\Phi}{\cos(\beta) \cos(\theta)}|\right]^\frac{1}{2}|\psi|$ in the 
compact set $|\lambda_i|\le 3\epsilon$ which is 
assured to be finite for every smooth function of rapid decrease.
Thus, we can approximate every $\psi \in \mathcal{S}$ through some 
$\psi_\epsilon \in \FC$ to arbitrary precision. Hence, $A^n$ with domain 
$\FC$ is a densely defined operator in $L_2(\RR^9)$ and therefore all 
the operators defined in the last section are actually well defined and 
even more they leave the domain invariant.

\section{Coherent states for the gravitational sector} 
\label{sec:classical_limit}

In this section we want to construct coherent states for the 
gravitational sector: The gravitational Hilbert space introduced in 
section \ref{Hspace} is at each vertex of the algebraic graph given by 
${\cal H}_v^g=L_2(\mathbb{R}^9, d\mu)$ and thus consists of nine copies 
of  usual Schr\"odinger representation Hilbert spaces.  We will 
construct coherent states for each vertex Hilbert space 
${\cal H}_v^g$ and the coherent state for the complete gravitational 
sector are then given by the infinite tensor product of the vertex 
coherent states. Since the quantum theory is formulated on a purely 
algebraic level, the coherent states will also be the bridge between the 
algebraic quantum theory and the embedded classical theory of General 
Relativity.\\
\\
In order to construct coherent states we first need to choose an 
embedding $X$ of our algebraic graph $\alpha$ in to a given manifold 
$\Sigma$ and we call its image $\gamma:=X(\alpha)$.
We will choose embeddings $X$ such that $\gamma$ is dual to a certain 
triangulation $\gamma^ *$ . Thus, for each embedded edge $X (c)$ 
there is a face $S_c$ in $\gamma^*$ which intersects $X(c)$ only in an 
interior point $p_c$ of both $S_c$ and $X (c)$. Next we need to choose a 
classical cotriad $E_0$ (suppressing indices) and
an $su(2)$ -- valued vector density $P_0$. With the manifold, the 
embedding and the classical data $(E_0,P_0)$ we can define the smeared 
quantities
\be
{E^i_I}_0(v)&=&\int\limits_{X(c_I)} E_0\nonumber\\ 
{P^I_i}_0(v)&=&\int\limits_{S_{c_I}}  \epsilon_{abc} d\sigma^a\wedge 
d\sigma^b P^c_i(\sigma)
\ee
 We follow the complexifier method for coherent states introduced in 
\cite{GCSIV,GCSI,GCSII,GCSIII}. 
We complexify the configuration space $\RR^9$ to $\mathbb{C}^9$ by 
introducing 
\be
Z^{j}_J(v):=\frac{1}{\ell_p\sqrt{2}}\left({E^j_J}_0(v) - i 
{P^J_j}_0(v)\right)
\ee
From now on we will suppress indices and denote the classical data 
simply by $Z(v)=(E_0(v),P_0(v))$.
This complexification can be obtained from a classical complexifier of 
the form
\be
C&:=&\frac{1}{2}\frac{1}{\kappa\ell_p\sqrt{2}}\sum\limits_{v\in 
V(\alpha)}P^{J}_j(v)P_J^j(v)
\ee
with
\be
Z^{j}_J(v)=\sum\limits_{n=0}^\infty \frac{i}{n!}\{E^j_J(v),C\}_{(n)}
\ee
where $\{.,.\}_{(n)}$ denotes the iterative Poisson bracket defined as 
$\{E^j_J(v),f\}_{(0)}=E^j_J(v)$ and  then iteratively by\\ 
$\{E^j_J(v),f\}_{(n+1)}=\{E^j_J(v),\{E^j_J(v),f\}_{(n)}\}$.
Similarly we can define a corresponding complexifer for the quantum 
configuration space 
\be
\hat{C}&:=&\frac{1}{2}\frac{1}{\ell_p^2\sqrt{2}}\sum\limits_{v\in 
V(\alpha)}\left(i\ell_p^2\frac{\partial}{\partial{E^j_J}(v)}\right)\left(i\ell_p^2\frac{\partial}{\partial{E^k_K}(v)}\right)\delta_{jk}\delta^{JK}
\ee
from which we obtain the form of the annihilation  and creation 
operators 
\be
A^{j}_J(v)&:=&\frac{1}{\ell_p\sqrt{2}}\left({E^i_I}(v) - i 
\left(i\ell_p^2\frac{\partial}{\partial{E^{j_J}(v)}}\right)\right)\\
(A^{j}_J)^{\dagger}(v)&:=&\frac{1}{\ell_p\sqrt{2}}\left({E^i_I}(v) + i 
\left(i\ell_p^2\frac{\partial}{\partial{E^{j_J}}(v)}\right)\right)
\ee
that satisfy the algebra 
$[A^{j}_J(v),(A^{k}_K)^{\dagger}(v')]=\delta_{vv'}\delta^{jk}\delta_{JK}$. 
Note that the characteristic length that enters into the definition of 
the coherent states is $\ell_p$ in our case and not as for the harmonic 
oscillator  $\ell=\sqrt{\hbar/m\omega}$.  {The reason for this is simply 
that the squared Planck length $\ell_p^2$ also occurs  in the definition 
of the momentum operator.}
One of the defining properties of coherent states is that they are 
eigenstates of the annihilation operator $A^j_J$ with eigenvalue $Z$, 
the classical phase space point that they are labeled by and 
around which they are sharply peaked. They can be expressed as
\be
|\psi_Z\rangle &:= &\sum\limits_{n=0}^\infty 
\exp(-\frac{1}{2}|Z|^2)\frac{Z^n}{n!}\Big[\left(A^{j}_J(v)\right)^{\dagger}\Big]^n 
| 0\rangle
=\exp\left(-\frac{1}{2}|Z|^2\right)\exp\left(Z(A^{j}_J)^{\dagger}(v)\right)| 
0\rangle
\ee
where $| 0\rangle$ is the vacuum state defined by $A^{j}_J(v)| 0\rangle 
=0$. Further properties of coherent states are that they satisfy a 
resolution of identity, they form an overcomplete basis, and that they 
are not mutually orthogonal but $\langle\psi_{Z^{\prime}}\, ,\, 
\psi_Z\rangle = \exp(-|Z - Z^{\prime}|^2)$.
\\
In position space representation we get using $Z(v)=(E_0(v),P_0(v))$
\be
\psi_Z(E^j_J(v))&=&\frac{1}{\sqrt{\ell_p\sqrt{\pi}}}\exp\left(-\frac{i}{2\ell_p^2}P_0(v)E_0(v) 
+\frac{i}{\ell_p^2}P_0(v)E^j_J(v)\right)
\exp\left(-\frac{1}{2\ell_p^2}\left(E^j_J(v)-E_0(v)\right)^2\right)
\ee
 An easy calculation shows that $\psi_Z(E^j_J)$ is indeed an eigenstate 
of $A^{j}_J(v)$ with eigenvalue $Z(v)$.
 The states $\psi_Z(E^j_J)$ are the ordinary harmonic oscillator 
coherent states and thus we can construct the 
coherent states associated to each vertex as a product of nine harmonic 
oscillator coherent states and the total states for the gravitational 
sector then as infinite product of states over all vertices  {of the} 
graph
\be
\psi_Z(v)&:=&\prod\limits_{j,J=1}^3\psi_Z(E^j_J(v))\in{\cal 
H}_v^g\quad\rm{and}\quad \psi_Z:=(\otimes_v \psi_Z(v))\in{\cal 
H}_{\otimes}^g
\ee
 {The fact that $\psi_Z$ is an eigenstate of the annihilation operator 
$A^{j}_J$ with eigenvalue $Z$ has the consequence that it approximates 
expectation values of $E^j_J$ and $P^J_j$ semiclassically
well. Precisely, we have}
\be
\langle\psi_Z\, ,\, E^j_J\,\psi_Z\rangle &=& E_0\, ,\quad
\langle\psi_Z\, ,\, P^J_j\,\psi_Z\rangle = P_0
\ee
However, when we consider the  inverse operator $1/\det(E^j_J)$ it turns 
out its expectation value with respect to $\Psi_Z$ diverges. A simple way 
to see why this happens is to take a cotriad configuration at one vertex 
where all but two triads are vanishing. Then we have at least one 
integral of the symbolic form $\int_\RR e^{-x^2}/(x-x_0)dx$ 
which 
does not converge. For this reason we have to introduce modified 
coherent states along the lines of the modified Schwartz functions of 
rapid decrease in section \ref{sec:operators}. Thus we define the 
modified coherent states by means of the function $s_\epsilon$ in 
equation (\ref{seps}) as
\be
\psi_Z^{\epsilon}(v)&:=&\prod\limits_{j,J=1}^3s_\epsilon(\det(E^j_J(v))\psi_Z(E^j_J(v))\quad\rm{and}\quad\psi^\epsilon_Z:=(\otimes_v 
\psi^\epsilon_Z(v))
\ee
More precisely
\be
\psi^\epsilon_Z(v): \RR^9 \rightarrow \RR; \quad \Omega, \lambda_i 
\mapsto 
\psi^\epsilon_Z(\Omega, \lambda_i) = s_\epsilon(\lambda_1) 
s_\epsilon(\lambda_2) s_\epsilon(\lambda_2) \psi_Z(\Omega, \lambda_i)
\ee 
Using the same steps as in equation \ref{Psiest} we obtain
\be
||\psi_Z^{\epsilon} - \psi_Z||
& \leq & \IC_{\rm max} \sqrt{6^3\epsilon^3} \left(\int d\Omega 
|\cos(\beta) \cos(\theta)| \right)^{\frac{1}{2}} \propto 
\epsilon^{\frac{3}{2}} \quad
\ee
where here $\IC_{\rm max}$ is the supremum of $\left[ \frac{\det 
\Phi}{\cos(\beta) \cos(\theta)}\right]^\frac{1}{2}|\psi_Z|$ in 
$|\lambda_i|\le 3\epsilon$ which is 
finite for smooth functions.
Hence, for sufficiently small $\epsilon$ we can approximate the states 
$\psi_Z$ by the states $\psi_Z^\epsilon$ for all values of the classical 
phase space point label $Z$. Furthermore, because the coherent states 
have the property of being sharply peaked around $Z$, the absolute 
values 
of $\psi_Z$ that appears in $ \IC_{\rm max}$ in the estimate above 
becomes smaller and smaller the farther  the classical phase space point 
$Z$ is away from a singular configuration. Consequently as long as we 
are mainly interested in states which are not peaked around 
singularities the approximation works very well. Indeed we are actually 
interested in those states because we want to use these states to mimic 
a classical geometry background when investigating the 
 {semiclassical limit  of this theory understood as QFT on curved 
spacetimes in section \ref{sec:analysis}.} 

We will now verify that the expectation 
values for $E^j_J$ and $P^J_j$ are also well approximated for 
sufficiently small $\epsilon$.
The modified states are no longer eigenstates of $A^j_J$  with 
eigenvalue $Z$ but we get  a  correction involving the derivative of 
$s_\epsilon$ and the original state $\psi_Z$.
\be
A^j_J\psi^\epsilon_Z &= &Z \psi^\epsilon_Z 
+\frac{\ell_p}{\sqrt{2}}\frac{\partial s_\epsilon(\det(E^j_J)}{\partial 
E^j_J}\psi_Z
\ee
This correction term only appears when $|\det(E^j_J)|\leq\epsilon$ since 
otherwise $s_\epsilon(\det(E^j_J))=1$. For the  error of the expectation 
value of $\hat{E}^j_J$ we therefore get
\be
\lefteqn{\Big|\langle \psi_Z^\epsilon,\, \hat{E}^j_J\, 
\psi^\epsilon_Z\rangle - \langle \psi_Z \hat{E}^j_J\, \psi_Z\rangle\Big|
=
\Big|\langle \left(s_\epsilon +1\right)\psi_Z,\, \hat{E}^j_J\, 
\left(s_\epsilon -1\right)\psi_Z\rangle\Big|} \nonumber\\
&=&
\Big|\int\limits_{\RR^9} d^9E^j_J  \left(s_\epsilon(\det(E^j_J)) 
+1\right)\bar{\psi}_Z(E^j_J)E^j_J\left(s_\epsilon(\det(E^j_J)) 
-1\right)\psi_Z(E^j_J)\Big| \nonumber\\
&=&
\Big|\int\limits_{\RR^9} d\Omega d^3\lambda\det(\Phi) 
\left((s_\epsilon(\lambda_1)s_\epsilon(\lambda_2)s_\epsilon(\lambda_2))^2 
-1\right)\bar{\psi}_Z(\lambda_i,\Omega)E^j_J(\lambda_i,\Omega) 
\psi_Z(\lambda_i,\Omega)\Big| \nonumber\\
&=&
\int\limits_{[S_3/Z_2] \times [S_3/Z_2]} 
d\Omega\int\limits_{-3\epsilon}^{3\epsilon}d\lambda_1\int\limits_{-3\epsilon}^{3\epsilon}d\lambda_2\int\limits_{-3\epsilon}^{3\epsilon}d\lambda_3\big|\det(\Phi)\big|
\big| 
\left((s_\epsilon(\lambda_1)s_\epsilon(\lambda_2)s_\epsilon(\lambda_2))^2 
-1\right)\big|\big|\bar{\psi}_Z(\lambda_i,\Omega)E^j_J(\lambda_i,\Omega) 
\psi_Z(\lambda_i,\Omega)\big| \nonumber\\
&\leq&
\int\limits_{[S_3/Z_2]  \times [S_3/Z_2]} 
d\Omega\int\limits_{-3\epsilon}^{3\epsilon}d\lambda_1\int\limits_{-3\epsilon}^{3\epsilon}d\lambda_2\int\limits_{-3\epsilon}^{3\epsilon}d\lambda_3
\big|\det(\Phi)\big| 
\big|\bar{\psi}_Z(\lambda_i,\Omega)E^j_J(\lambda_i,\Omega) 
\psi_Z(\lambda_i,\Omega)\big| \nonumber\\
&\leq&
  \IC_{\rm max}6^3\epsilon^3\left(\int d\Omega\; |\cos(\beta) 
\cos(\theta)| 
\right)\propto \epsilon^3
\ee
with  $\IC_{\rm max}$ is the supremum of $\left[ \frac{\det 
\Phi}{\cos(\beta) 
\cos(\theta)}\right]^\frac{1}{2}|\bar{\psi}_ZE^{j}_{j}(\lambda_i)\psi_Z|$
in $|\lambda_i|\le 3\epsilon$. 
We used that $E,s_\epsilon$ commute and 
in step three we performed again a transformation to the 
$(\lambda,\Omega)$ coordinates and considered $E^j_J$ as a function of 
these variables.
Similarly we obtain for the momentum operator
\be
\Big|\langle \psi_Z^\epsilon,\, \hat{P}^J_j\, \psi^\epsilon_Z\rangle - 
\langle \psi_Z, \hat{P}^J_j\, \psi_Z\rangle\Big|
&=&
\Big|\langle [\psi_Z^\epsilon-\psi_Z],\, \hat{P}^J_j\, 
\psi^\epsilon_Z\rangle + 
\langle \psi_Z, \hat{P}^J_j\, [\psi^\epsilon_Z-\psi_Z]\rangle\Big|
\\
&=&
\Big|\langle [\psi_Z^\epsilon-\psi_Z],\, [[\hat{P}^J_j s_\epsilon]\; 
\psi_Z+s_\epsilon\;[\hat{P}^J_j \psi_Z]]\rangle + 
\langle  \hat{P}^J_j\psi_Z, \, [\psi^\epsilon_Z-\psi_Z]\rangle\Big|
\nonumber\\
&=&
\Big|\langle \psi_Z,\, \frac{1}{2}[[\hat{P}^J_j (s_\epsilon-1)^2]\; 
\psi_Z+[s_\epsilon-1]\;s_\epsilon\;[\hat{P}^J_j \psi_Z]]\rangle + 
\langle  \hat{P}^J_j\psi_Z, \, [s_\epsilon-1]\;\psi_Z\rangle\Big|
\nonumber\\
&\leq&
\int\limits_{\RR^9} d\Omega d^3\lambda\big|\det(\Phi)\big|\; 
\left(s_\epsilon(\lambda_1)s_\epsilon(\lambda_2)s_\epsilon(\lambda_2) 
-1\right)^2\;\big|i\ell_p^2\frac{\partial}{\partial 
E^j_J(\lambda_i,\Omega)} |\psi_Z(\lambda_i,\Omega)|^2\big| \nonumber\\
&& +\Big|\int\limits_{\RR^9} d\Omega 
d^3\lambda\;|\det(\Phi)|\;|\bar{\psi}_Z(\lambda_i,\Omega)|\;
|\frac{\partial}{\partial E^j_J(\lambda_i,\Omega)} 
\psi_Z(\lambda_i,\Omega)|\;
\Big|s_\epsilon(\lambda_1)s_\epsilon(\lambda_2)s_\epsilon(\lambda_2)-1
\Big|
\ee
where we introduced the abbreviation $\partial 
E^j_J(\lambda_i,\Omega)=(\partial\lambda_i/\partial 
E^j_J)\partial\lambda_i+(\partial\Omega_i/\partial 
E^j_J)\partial\Omega_i$ for the transformed partial derivative. Here 
in the last step we have performed an integration by parts. 
As before, both terms above are only non -- 
vanishing in the compact interval $\lambda_i\in [-3\epsilon, 
3\epsilon]$. Restricting the integration domain and using the upper 
bounds $|s_\epsilon|,\;|1-s_\epsilon|\le 1$ we see that both terms are 
bounded 
by $[6\epsilon]^3\;[\int d\Omega\;|\cos(\beta)\cos(\theta)|]$ times the 
supremum of the respective integrand on the domain $|\lambda_i|\le 
3\epsilon$. Since that supremum does not increase if we lower $\epsilon$ 
we see that the bound can be arbitrarily small.
Thus also for the expectation value of $P_j^J$ we are arbitrarily close 
to the original value when using the modified coherent states.
Finally, let us compute the expectation value for the operator 
$A^n:=[1/\det(E^j_J)]^n$ with respect to the modified states. To display 
the exact formulas requires a lot of notation, so we satisfy ourselves 
with giving the flavour of it by considering a simpler example which 
nevertheless contains all the essential features. The exact calculation 
proceeds completely analogously. We model the expectation value of $A^n$
by 
\be \label{1.1}
<\frac{1}{|.|^n}>:=
\int_{\mathbb{R}}\;\frac{dx}{\sqrt{\pi}\delta}\;e^{-(x-x_0)^2/\delta^2}\;
\frac{s_\epsilon(x)^2}{|x|^n}
\ee
and want to compare its value with $1/|x_0|^n$. We write
\be \label{1.2}
<\frac{1}{|.|^n}>:=I_1+I_2:=
\int_{\epsilon\le |x|\le 3\epsilon}\;
\frac{dx}{\sqrt{\pi}\delta}\;e^{-(x-x_0)^2/\delta^2}\;
\frac{s_\epsilon(x)^2}{|x|^n}
+\int_{|x|\ge 3\epsilon}\;
\frac{dx}{\sqrt{\pi}\delta}\;e^{-(x-x_0)^2/\delta^2}\;
\frac{1}{|x|^n}
\ee
and have due to $|s_\epsilon| \le 1$ 
\be \label{1.3}
|I_1| &\le &
\int_{\epsilon\le |x|\le 3\epsilon}\;
\frac{dx}{\sqrt{\pi}\delta}\;e^{-(x-x_0)^2/\delta^2}\;
\frac{1}{|x|^n}
\nonumber\\
&\le&
\frac{1}{\epsilon^n}\int_\epsilon^{3\epsilon}\;
\frac{dx}{\sqrt{\pi}\delta}\;
[e^{-(x-|x_0|)^2/\delta^2}+e^{-(x+|x_0|)^2\delta^2}]
\nonumber\\ & \le &
\frac{2}{\epsilon^{n-1}\sqrt{\pi}\delta}\;
[e^{-(|x_0|-3\epsilon)^2/\delta^2}+e^{-(|x_0|+\epsilon)^2/\delta^2}]
\ee 
where we have assumed that $|x_0|\gg \epsilon$. Next
\be \label{1.4}
I_2-\frac{1}{|x_0|^n}:=-I_3+I_4:=
-\int_{|x|\le 3\epsilon}\;
\frac{dx}{\sqrt{\pi}\delta}\;e^{-(x-x_0)^2/\delta^2}\;
\frac{1}{|x_0|^n}
+\int_{|x|\ge 3\epsilon}\;
\frac{dx}{\sqrt{\pi}\delta}\;e^{-(x-|x_0|)^2/\delta^2}\;
\left[\frac{1}{|x|^n}-\frac{1}{|x_0|^n}\right]
\ee
where in the last integral we have exploited the invariance under $x\to 
-x$.
We can estimate 
\be \label{1.5}
|I_3|\le 
\frac{3\epsilon}{\sqrt{\pi}\delta |x_0|^n}\;
[e^{-(|x_0|-3\epsilon)^2/\delta^2}+e^{-(|x_0|)^2/\delta^2}]
\ee
To estimate $I_4$ consider 
\be \label{1.6}
f(x):=-\frac{(x-|x_0|)^2}{\delta^2}-n\ln(|x|)
\ee
The extremum $y$ of $f$ satisfies $2x(x-|x_0|)=-n\delta^2$ i.e.
$y_\pm=[|x_0|\pm\sqrt{x_0^2-n\delta^2}]/2$ assuming $|x_0|\gg \delta$. 
Hence $y_+\approx |x_0|,\;y_-\approx n\delta^2/4|x_0|$. Since 
$f^{\prime\prime}(x)=-2/\delta^2+n/x^2$ we find 
\be \label{1.7}
f^{\prime\prime}(y_+)\approx -2/\delta^2+n/x_0^2<0,\;\;
f^{\prime\prime}(y_-)\approx 1/\delta^4[-2\delta^2+16x_0^2/n]>0
\ee
Thus only $y_+$ is a maximum and it also satisfies $|y_+|\ge 3\epsilon$. 
Accordingly, a saddle point estimate of $I_4$ yields
\be \label{1.8}
|I_4| 
&\approx& \left|2\int_{\epsilon}^\infty\; 
\frac{dx}{\sqrt{\pi}\delta}\;\left[e^{f(y_+)+f^{\prime\prime}(y_+)[x-y_+]^2}-
e^{-(x-|x_0|)^2/\delta^2}/|x_0|^n\right]\;\right|
\nonumber\\
&\approx& 2\left|\int_{\epsilon}^\infty\; 
\frac{dx}{\sqrt{\pi}\delta}\;e^{-(x-|x_0|)^2/\delta^2}/|x_0|^n\;
\left(\left[\frac{|x_0|}{y_+}\right]^n-1\right)\right|
\nonumber\\
&\le& 
\left|\frac{1}{|x_0|^n} \left(\left[\frac{|x_0|}{y_+}\right]^n-1\right)\right|\approx 
\frac{1}{|x_0|^n} \; \frac{n^2 \delta^2}{4|x_0|^2}
\ee
Summarising
\be \label{1.9}
\left|<\frac{1}{|.|^n}>|x_0|^n-1\right|\le (|I_1|+|I_3|+|I_4|)\;|x_0|^n\;
\le
\frac{n^2\delta^2}{4|x_0|^2}+
\frac{6\epsilon}{\sqrt{\pi}\delta}\;e^{-|x_0|^2/\delta^2}+
\frac{4|x_0|^n}{\epsilon^{n-1}\sqrt{\pi}\delta}\;
e^{-|x_0|^2/\delta^2}
\ee
In our application $\delta=\ell_P$ is fixed while $\epsilon\to 0$. 
We have already seen that $|x_0|\gg \epsilon,\delta$ in order that 
(\ref{1.9}) is a good estimate. Let $y=|x_0|/\delta$ then the third term 
in (\ref{1.9}) is of the form
\be \label{1.10}
\left[\frac{\epsilon}{\delta}\right]^{n-1}\; y^n\; e^{-y^2/2}
\ee
which should be small as compared to the fluctuation term $n^2 
\delta^2/4|x_0|^2$. This shows that we cannot let $\epsilon\to 0$ for our 
approximate coherent states. However, choosing e.g. $\epsilon=\delta$ we 
see that for $n$ of the order unity already for $y$ of the 
order $\sqrt{n}$ the third term in (\ref{1.9}) is subdominant.

\section{Analysis of the system} \label{sec:analysis}

\subsection{Born--Oppenheimer approximation}

To analyse the interplay between gravitational and matter dynamics and 
to understand how a (matter--) quantum field theory on a fixed classical 
spacetime can emerge from the fully quantum gravitational setup we 
want to employ an approximation scheme of Born--Oppenheimer type. As we 
will see in the following for such an approximation scheme to work it is 
essential that the geometrical variables are quantised as multiplication 
operators, not as derivative operators. This is the main motivation for 
not using the LQG--representation but the one that we constructed in 
previous sections. This representation captures some of the features of 
LQG (or the recently proposed generalisation coined Algebraic Quantum 
Gravity \cite{AQGI, AQGII, AQGIII}), especially the fundamentally 
discrete graph--like structures which appear in the very definition of 
the kinematical Hilbert space. In order to be able to use a 
Born--Oppenheimer approximation scheme we have chosen a representation 
of the gravitational Poisson algebra which is commutative in the 
geometrical variables, unlike the non--commutative flux operators in  
{ordinary} LQG. Thus, the theory constructed in this paper is a close 
relative to LQG when defined on an algebraic graph which shares some of 
its features 
(based on graph--like structures) but deviates from it in other 
essential characteristics (abelian geometrical operators). Of course, 
the long term goal will be to generalise the methods derived in this 
section to full LQG also taking into account the non--commutative 
structure of the holonomy--flux algebra. But, as we will see below, due 
to the specific assumptions made in the derivation of the 
Born--Oppenheimer approximation such a generalisation is not 
straightforward and will require substantially more work. One step into 
that direction is the construction of a non--commutative flux 
representation for the holonomy--flux algebra, where the flux--operators 
act as multiplication operators on an appropriately defined Hilbert space 
and the non--abelian nature of the flux--operators is taken into account 
through a specific $*$--product on that space 
\cite{flux_representation}.\\
The use of Born--Oppenheimer approximation schemes in the context of 
quantum gravity has a long tradition, especially in the older 
Wheeler-deWitt approach (for a historical account see for example 
Kiefer's book \cite{kiefer_book} and references therein, for a recent 
application in the framework of spinfoam models see 
\cite{rovelli_vidotto}). Most of the results so far were obtained on a 
rather formal level, because the Hilbert space on which Wheeler-deWitt 
theory is assumed to be defined is not known at all, but the application 
to minisuperspace models for FRW--universes gives reasonable results and 
can shed some light on conceptual questions such as the often discussed 
problem of time.\\
Concerning this issue we are in a better position: The Hilbert space of 
the quantum gravity theory we consider is explicitly known and all the 
operators we use are densely defined symmetric operators on that space. 
Therefore the derivations which follow are all well defined and work 
beyond the formal level -- at least in principle when neglecting 
practical difficulties in actually performing the calculations. 
Moreover, we are advocating a reduced phase space quantisation of  
{General Relativity} where the preferred reference frame is given by 
Brown--Kuch\v{a}r dust fields. Thus we have a true physical Hamiltonian 
that generates evolution in dust--time, not only a vanishing Hamiltonian 
constraint. The problem of time is therefore solved already on the 
classical level and does not import any extra obstructions into the 
quantum theory.

\subsubsection{General Framework} For simplicity we will first explain 
the Born--Oppenheimer approximation using a very simple quantum 
mechanical system and later generalise to the quantum gravitational 
setup. Assume that our classical phase space is four dimensional and 
coordinatised by two pairs of canonically conjugate variables $(Q,P)$ 
and $(q,p)$  (a generalisation to an arbitrary  number of configuration 
variables is straightforward but will not be considered here for 
pedagogical reasons). Further, assume that dynamics is generated by a 
Hamiltonian of the form
\be
H := \frac{P^2}{2 M} + \frac{p^2}{2 m} + V(Q,q) \quad ,
\ee
where $M$ and $m$ are some parameters describing the system under 
investigation (in the case of two particles these will be their 
respective masses) and $V(Q,q)$ is a potential term which depends only 
on the configuration variables, not on their momenta.\\
This system can be quantised on a Hilbert space $\HC := \HC^Q \otimes 
\HC^q$ where $\HC^Q$ and $\HC^q$ are taken to be the usual spaces of 
square integrable functions over the real line. On $\HC$ one defines 
operators in the usual way as $(\hat{q}\Psi)(Q,q) := q \Psi(Q,q), 
(\hat{Q}\Psi)(Q,q) := Q \Psi(Q,q), (\hat{p}\Psi)(Q,q) := i\hbar 
\partial_q \Psi(Q,q), (\hat{P}\Psi)(Q,q) := i\hbar \partial_Q \Psi(Q,q)$ 
for functions $\Psi \in \HC$. Thus, the Hamilton operator of the total 
system is given by
\be
\hat{H} := -\frac{\hbar^2}{2M}\Delta_Q  -\frac{\hbar^2}{2m}\Delta_q + 
V(Q,q) \quad ,
\ee
where $\Delta_Q, \Delta_q$ denotes the Laplacian with respect to $Q$ and 
$q$ respectively.\\
In general, depending on the exact expression for the potential term 
$V(Q,q)$, it can be difficult to obtain solutions to the full eigenvalue 
problem
\be
\hat{H} \Psi^\alpha = \Lambda^\alpha \Psi^\alpha \quad , 
\label{Schroedinger_Qq}
\ee
where $\alpha$ is a set of quantum numbers labelling the eigenvalues 
$\Lambda^\alpha$ and the corresponding eigenfunctions $\Psi^\alpha$.
However, under certain assumptions it is possible to obtain approximate 
solutions: Assume that (i) $m \ll M$, which means that there are two 
clearly separated energy scales in the problem at hand, and (ii) we have 
some control over one half of the problem, namely we can solve the 
eigenvalue problem for the light variables
\be
\hat{\tilde{H}}(Q) \chi_i(q; Q) = \lambda_i(Q) \chi_i(q; Q) \quad , 
\label{Schroedinger_q}
\ee
where $\hat{\tilde{H}}(Q) := -\frac{\hbar^2}{2m}\Delta_q + V(Q,q)$ is an 
operator acting on the space $\HC^q$ that depends on an external 
parameter $Q$ and $\chi_i(q;Q)$ (labeled by a set of quantum numbers 
$i$) are assumed to be square integrable in $\HC^q$ and also depending 
on $Q$ in a parametrical way. So in a sense this means we can solve 
\ref{Schroedinger_q} for each $Q$ separately.\\
Let us first stick to the first half of the problem and analyse a little 
further equation (\ref{Schroedinger_q}): Assume that we can solve the 
eigenvalue problem for each value of $Q$ and we have obtained 
eigenfunctions $\chi(q; Q) \in \HC^q$ and eigenvalues $\lambda(Q)$ that 
depend on the external parameter $Q$. Then a priori, it is not clear how 
$\chi$ and $\lambda$ behave if we change $Q$\footnote{To get some 
intuition for this problem one could think of a quantum particle under 
the influence of an external magnetic field $Q$. Assume that we can 
solve the Schroedinger equation for each fixed value of the magnetic 
field. Then the question would be: How does the quantum system 
describing the particle change when we manually change the external 
magnetic field?}.
Berry \cite{berry} showed that this question can be answered in a nice 
geometric way: Under the assumption of adiabaticity\footnote{The 
argument turns out to work out in more general situations, see for 
example \cite{shapere_wilczek_book}.} (i.e. assuming that the change in 
$Q$ is so slow that the quantum system at time $t$ is in a state 
$\chi(q;Q(t))$) it turns out that when moving the quantum system around 
a closed curve $C$ in parameter space (in our simple example the 
parameter space would just be $\RR$ but the same phenomenon holds also 
for higher dimensional parameter spaces) the wave function $\chi$ picks 
up a non--trivial phase factor. That means when the system is prepared 
to be in an energy eigenstate $\chi_i(0)$ for $t=0$, after having 
followed the loop $C$ in a time interval $T$ it turns out to be in the 
state $\psi(T) = \exp\left[ \frac{i}{\hbar}\gamma_i(C)\right] \exp\left[ 
-\frac{i}{\hbar}\int \limits_0^Tdt \lambda_i(Q(t))\right]\chi_i(0)$. The 
second exponential is just the usual dynamical phase factor, but the 
first one, explicitly given by
\be
\gamma_i(C) = i\hbar \oint\limits_C \braaaket{  \chi_i(Q)}{ \partial_Q}{ 
\chi_i(Q)}_q \quad ,
\ee
is indeed non--trivial. This phase factor, known as {\it Berry's phase} 
in the literature was experimentally confirmed in many experiments and 
can be used to explain a variety of physical phenomena, including for 
example the Aharonov--Bohm effect (see \cite{shapere_wilczek_book} for 
more information on this broad topic). From a geometrical point of view 
the situation is quite interesting (see for example 
\cite{geometric_phases_book} and references in there for a modern 
explanation using the language of fiber bundles), because it turns out 
that $A := i\hbar \braaaket{ \chi_i}{ \partial_Q}{ \chi_i}_q$ transforms 
as a $U(1)$--connection under coordinate transformations in parameter 
space. Generalised to a higher--dimensional parameter space this 
connection is given by $A_\mu := i\hbar \braaaket{ \chi_i}{ 
\frac{\partial}{\partial Q^\mu} }{ \chi_i }_q$ in a local coordinate 
system $Q^\mu$. Thus, $\exp\left[ \frac{i}{\hbar}\gamma_i(C)\right]$ is 
nothing else than the holonomy of a $U(1)$--connection along the path 
$C$ in parameter space.\\
One essential point in the derivation of the above formulas is that $Q$ 
is treated as an external classical parameter: To be able to talk about 
a path in parameter--space one must assume that $Q$ is a classical 
variable so that it makes sense to demand differentiability of the 
paths. 
Strictly speaking, this assumption is violated when we take into account 
the quantum nature of $Q$ itself, because in quantum theory particles do 
not follow differentiable paths anymore\footnote{In fact, the path 
integral measure is concentrated on paths that are arbitrarily 
discontinuous.}, and taking this quantum nature into account is a highly 
non--trivial task (see again \cite{shapere_wilczek_book} and especially 
\cite{moody_shapere_wilczek} for elaborations on this topic). However, 
as long as we are interested in a system where the respective energy 
scales are well separated ($m \ll M$) and we are working in a 
representation where $\hat{Q}$ acts as a multiplication operator and not  
{as} a derivative operator the approximation seems to be valid and is 
confirmed by many experiments  {in molecular physics}.\vspace{.2cm}\\
Now that we understood the first half of the problem, let us go back to 
the full problem (\ref{Schroedinger_Qq}) and see how we can obtain 
solutions to the eigenvalue problem for the full Hamilton operator 
$\hat{H}$ acting on the fullg space $\HC$:
Assuming that we have solved the eigenvalue problem for 
$\hat{\tilde{H}}(Q)$ with $Q$ as an external parameter and we already 
know its eigenfunctions $\chi_i(q;Q)$ we start with the Ansatz
\be
\Psi^\alpha(Q,q) := \sum\limits_{i}\xi_i^\alpha(Q)\chi_i(q; Q) \quad . 
\label{ansatz_bo}
\ee
Plugging this into (\ref{Schroedinger_Qq}) leads to
\be
\hat{H} \Psi^\alpha = \sum\limits_i(-\frac{\hbar^2}{2M}\Delta_Q +  
\lambda_i(Q) ) \xi_i^\alpha(Q) \chi_i(q; Q)  = \Lambda^\alpha 
\sum\limits_i \xi_i^\alpha(Q) \chi_i(Q; q)\quad .  
\ee
Now we multiply by $\chi_k$ from the left and take the scalar product in 
$\HC^q$ to receive an equation for the Q--dependent coefficients 
$\xi_i^\alpha$
\be
\sum\limits_i \left(  \braaaket{  \chi_k}{ 
\left(-\frac{\hbar^2}{2M}\Delta_Q\right)}{ \chi_i   }_q  + 
\lambda_i(Q)\delta_{ik} \right) \xi_i^\alpha  = \Lambda^\alpha 
\xi_k^\alpha
\ee
where both $\chi$ and $\xi$ are functions of $Q$ and the Laplacian 
$\Delta_Q$ acts on everything on its right. This can be rewritten as
\be
\sum\limits_i \left(\sum\limits_l  \frac{1}{2M}(-i\hbar \delta_{kl} 
\partial_Q -  A_{kl})(-i\hbar \delta_{li}\partial_Q -  A_{li} )+ 
\lambda_i(Q)\delta_{ik} \right) \xi_i^\alpha = \Lambda^\alpha 
\xi_k^\alpha
\ee
when using the connection
\be
A_{kl} := i\hbar \braaaket{  \chi_k}{  \partial_Q }{  \chi_l }_q \quad .
\ee
Thus we see that the effect of the light variables on the 
quantum--dynamics of the heavy system is twofold: First, there is an 
effective potential $\lambda_i(Q)\delta_{ik}$ which can be interpreted 
as the heavy variables experiencing the presence of the light variables 
only through an average. Second, the light variables effectively curve 
the space as seen by the heavy variables, because the ordinary momentum 
operator $\hat{P}_{kl} = -i\hbar \delta_{kl} \partial_Q $ gets replaced 
by the covariant momentum operator $\hat{\tilde{P}}_{kl} := 
-i\hbar(\delta_{kl} \partial_Q - \frac{i}{\hbar}A_{kl})$. Using the 
covariant momentum operators and defining $\hat{\tilde{P}}^2_{ik} := 
\sum\limits_l\hat{\tilde{P}}_{il}\hat{\tilde{P}}_{lk}$ the effective 
Schroedinger equation for the $Q$--dependent coefficients reads
\be
\sum\limits_i \left[\frac{\hat{\tilde{P}}_{ik}^2}{2M} + 
\lambda_i(Q)\delta_{ik}\right]\xi_i^\alpha = \Lambda^\alpha\xi_k^\alpha 
\quad . \label{Schroedinger_Q}
\ee
So far, apart from the assumption that $Q$ can be treated as a classical 
external parameter (i.e. showing no quantum behaviour as long as the 
dynamics of the $q$--variables is concerned), everything was exact. The 
next step would be to solve (\ref{Schroedinger_Q}) and thus to compute 
the coefficients $\xi_i^\alpha$ in the ansatz (\ref{ansatz_bo}). Thus we 
would arrive at a complete solution $\Psi^\alpha(Q,q)$ of the quantum 
system.\\
However, for general potentials $V(Q,q)$ (\ref{Schroedinger_Q}) turns 
out to be too complicated to be solved exactly, thus one has to rely on 
perturbation theory. The easiest route one can follow, which was 
originally proposed by Born and Oppenheimer in the context of molecule 
physics \cite{born_oppenheimer} and carries their name in the literature 
is to simply approximate
\be
\hat{\tilde{P}}_{ik} \approx \hat{P}\delta_{ik}
\ee
which diagonalises the operator in (\ref{Schroedinger_Q}) and leads to 
simply
\be
\left[\frac{\hat{P}^2}{2M} + \lambda_k(Q)\right]\xi_k^\alpha = 
\Lambda^\alpha \xi_k^\alpha \quad .
\ee
This means that the influence of the $q$--variables onto the quantum 
dynamics in the $Q$--sector is just taken into account via the effective 
potential term $\lambda_k(Q)$.
One can show that this approximation is justified whenever $|\lambda_i - 
\lambda_j| \gg |\Lambda^\alpha  -\Lambda^\beta|$ (see for example 
\cite{moody_shapere_wilczek}) and better approximations can be obtained 
by taking into account off-diagonal matrix elements or not setting the 
connection $A_{kl}$ to zero.\\
As we said before for the Born--Oppenheimer approximation to work one 
must assume that the $Q$--variables can be treated classically when we 
are only interested in the quantum dynamics of the $q$--system. This 
assumption is surely justified when working in a representation where 
$Q$ acts as a multiplication operator and furthermore the characteristic 
energy scales are well separated ($|\lambda_i - \lambda_j| \gg 
|\Lambda^\alpha  -\Lambda^\beta|$): To illustrate such a situation with 
a well known example consider a molecule with a number of nuclei with 
masses $M$ as $Q$--variables and a number of electrons with masses $m$ 
as $q$--variables. Semiclassically speaking, the electrons move much 
faster than the nuclei, therefore when treating the dynamics in the 
electron--sector one can safely treat the slow nuclei as classical 
variables.\\
There have been some efforts to generalise this picture and take into 
account the quantum nature of the $Q$--variables when treating the 
dynamics in the $q$--sector (see \cite{shapere_wilczek_book}) but the 
methods described above do not easily generalise to this setting. At 
least to our knowledge there are no results concerning the applicability 
of the Born--Oppenheimer method for non--commuting $Q$--variables and a 
generalisation into that direction seems to require some substantially 
new input.\\
The difficulties with the Born--Oppenheimer method for non--commuting 
variables was one of the main reasons to not choose the 
Ashtekar--Lewandowski--representation for gravity but 
to work with the one we constructed in this article. Here our 
geometrical 
variables (given by the cotriads) are represented as ordinary 
multiplication operators and thus the Born--Oppenheimer approximation is 
directly applicable.

\subsubsection{Application to gravity}
We want to employ an approximation scheme of the type described above to 
analyse the quantum theory for gravity plus matter which we have 
constructed in section \ref{sec:quantum}. Matter couples to gravity only 
via the cotriads, not via their canonically conjugate momenta, and due 
to the fact that we have chosen to quantise the cotriads as commuting 
multiplication operators a Born--Oppenheimer approximation scheme can 
directly be applied.
\vspace{.2cm}\\
We note that the system gravity plus matter has two widely separated 
energy scales: The {\it matter part} of the Hamiltonian carries 
information about the mass of the scalar field, which defines a length 
scale $\ell_\Phi := \frac{\hbar}{m}$. On the other hand, the Planck 
length $\ell_P$ enters the definition of the {\it gravitational part} of 
the Hamiltonian. For typical values of $m$ the quotient 
$\frac{\ell_P}{\ell_\Phi} \ll 1$. In fact this quotient will be much 
smaller than the quotient $\frac{m_{e^-}}{M_{\rm nuc}}$ in molecule 
physics, so one should expect that the Born--Oppenheimer approximation 
gives results of very good precision.\\
One can interpret this large separation of energy scales in a very 
intuitive way: During typical interaction processes between matter 
particles the geometry of spacetime changes very slowly. Of course, this 
will not be the case in the deep Planck regime, when considering 
processes with energies comparable to the Planck energy such that the 
quantum nature of spacetime itself has to be taken into account. But for 
all particle interactions in the semiclassical regime with centre of 
mass energies much lower than the Planck energy this should be a good 
approximation. Especially when considering a wave function that 
approximates a classical geometry (like the coherent states for the 
gravitational sector considered in section \ref{sec:classical_limit}) 
the influence of quantum (gravitational) fluctuations will be small. So 
in a sense this method allows us to compute the QFT on CS limit of 
quantum gravity. However, the method we will describe below allows us to 
go beyond this approximation and at least in principle compute quantum 
gravitational corrections to QFT on CS.
\vspace{.2cm}\\
Let us now explain the Born--Oppenheimer approximation adapted to our 
gravitational theory: The role of the heavy variables is played by the 
cotriads $E_I^i(v)$ and the role of the light variables is played by the 
matter fields $\Phi(v)$. To be more precise we are considering the 
Hilbert space $\mathcal{H}_\otimes =\otimes_v \mathcal{H}_v
=\otimes_v (\mathcal{H}_v^g \otimes \mathcal{H}_v^\phi)$, where 
 $\mathcal{H}^g_v = 
L_2(\mathbb{R}^9)$,  $\mathcal{H}^\phi_v = L_2(\mathbb{R})$. 
$\hat{E}_I^i(v)$ acts as a multiplication operator on 
$\mathcal{H}_\otimes$, $\hat{P}^I_i(v) := i \ell_P^2 
\partial_{E_I^i(v)}$ as a derivative operator thereon, $\hat{\Phi}(v)$ 
acts as a multiplication operator on $\mathcal{H}_\otimes$ and 
$\hat{\Pi}(v) = i\hbar \partial_{\Phi(v)}$ as a derivation operator 
thereon. The Hamiltonian is given approximately\footnote{In this 
approximation we keep only the first term in the power series $\bar{H} 
:= \hat{C}\left(   1  + \OC\left( \widehat{\frac{Q^{ab}C_a C_b}{C^2}} 
\right) \right)$ as explained earlier.} given by
\be
\hat{H}(\hat{P},\hat{E},\hat{\Pi},\hat{\Phi}) & \approx & \Big[ 
\hat{C}^g_{\rm kin}(\hat{P}, \hat{E}) + \hat{C}^g_{\rm pot}(\hat{E}) 
\Big] \otimes id_\Phi + \hat{C}^\phi(\hat{E}, \hat{\Pi}, \hat{\Phi}) 
\quad,
\ee
where $\hat{C}^\phi(\hat{E},\hat{\Pi}, \hat{\Phi}) =  \hat{C}^\phi_{\rm 
kin}(\hat{E},\hat{\Pi}) + \hat{C}^\phi_{\rm pot}(\hat{E},\hat{\Phi})$ 
factorizes into products of the form $\hat{O}^g \otimes \hat{O}^\phi$ 
with $\hat{O}^g \in \mathcal{L}(\mathcal{H}_\otimes^g)$ and 
$\hat{O}^\phi \in \mathcal{L}(\mathcal{H}_\otimes^\phi)$. One difference 
to the Hamiltonian in the toy model described above is that the kinetic 
terms $\hat{C}^g_{kin}$ and $\hat{C}^\phi_{\rm kin}$ are not functions 
of the momenta alone but are functions of the cotriad--operators 
$\hat{E}_I^i(v)$ as well. This does not spoil the general argument made 
above, however, the nice expressions in terms of a gauge potential 
$A_{kl}$ and its associated covariant derivative is not sufficient 
anymore and there will be extra terms.\\
We want to solve the eigenvalue problem
\be
\hat{H}(\hat{P},\hat{E}, \hat{\Pi}, \hat{\Phi}) \Psi^\alpha(E,\Phi) = 
\Lambda^\alpha \Psi^\alpha(E, \Phi) \quad , \label{schroedinger_full}
\ee
and for illustrative purposes we will assume that $\hat{H}$ has discrete 
spectrum such that $\Psi^\alpha(E, \Phi)$ are proper eigenfunctions and 
$\Lambda^\alpha$ their respective eigenvalues. At least as far as 
the matter part of the Hamiltonian is concerned, this will be the case 
whenever the spatial manifold of the continuum theory which arises as an 
approximation to the discrete theory is compact without boundary. The 
non--compact case has to be treated with more care and one needs to 
perform a direct integral decomposition of $\HC_\otimes$. However, these 
technical details will not be essential for our method and we will 
restrict ourselves to the case of compact spatial manifolds without 
boundary.\\
The strategy is exactly the same as in the toy model discussed before: 
We start with the Ansatz
\be
\Psi^\alpha(E,\Phi ) := \sum\limits_{i} \xi_i^\alpha(E)\chi_i(\Phi;E) 
\quad , \label{bo_ansatz}
\ee
for states $\Psi \in \HC_\otimes$ and assume that we already have some 
knowledge about the partial eigenvalue problem
\be
\hat{C}^\phi(\hat{\Phi}, \hat{\Pi}; E) \chi_i(\Phi;E) = \lambda_i(E) 
\chi_i(\Phi;E) \quad . \label{schroedinger_phi}
\ee
Here the dependence on $E$ has again to be understood in a parametric 
sense, i.e. we keep $E$ fixed and regard $\chi_i(\Phi; E)$ as an element 
of $\mathcal{H}_\otimes^\phi$. Thus, because $\hat{C}^\phi(\hat{\Phi}, 
\hat{\Pi}; E)$ is just the standard discrete matter Hamiltonian, this 
amounts to solving a certain type of lattice--QFT for an arbitrary (but 
fixed) background. Of course, in general this cannot be accomplished 
analytically and one will have to employ further approximation 
techniques, but the interesting point is that QFT on a curved spacetime 
emerges in the analysis of a theory of quantum gravity in the same way 
as the theory of quantised electrons in a given classical potential 
emerges out of the full quantum treatment of a molecule. For technical 
reasons one has to assume that  $\chi_i(\Phi;E)$ depends on $E$ in an at 
least twice differentiable manner.\\
In order to get the full wave functions $\Psi^\alpha(E, \Phi)$ we follow 
the Born--Oppenheimer approach as discussed for the toy model above, so 
we want to compute the coefficients $\xi_i^\alpha(E)$. This can be done 
by starting with (\ref{schroedinger_full}), multiplying by 
$\chi_k(\Phi;E)$ from the left and taking the scalar product in 
$\mathcal{H}_\otimes^\phi$. Assuming that $\chi_i(\Phi; E)$ is an 
orthonormal basis of $\mathcal{H}_\otimes^\phi$ we get
\be
\braaaket{ \chi_k(\Phi;E)}{ \hat{H} }{\Psi^\alpha(E, \Phi) }_\phi & = & 
\sum\limits_i \braaaket{ \chi_k(\Phi;E)}{ \Big[ \hat{C}^g_{\rm 
kin}(\hat{P}, \hat{E}) + \hat{C}^g_{\rm pot}(\hat{E}) \Big]}{ 
\chi_i(\Phi;E) }_\phi \xi_i^\alpha(E) + \lambda_k(E)\xi_k^\alpha(E) \nn 
\\
& = & \Lambda^\alpha\xi_k^\alpha(E) \label{eq1}
\ee
If we define $\hat{\mathbb{C}}^g_{ki} := \braaaket{ \chi_k(\Phi;E)}{ 
\Big[ \hat{C}^g_{\rm kin}(\hat{P}, \hat{E}) + \hat{C}^g_{\rm 
pot}(\hat{E}) \Big]}{ \chi_i(\Phi;E) }_\phi$  as an operator in 
$\mathcal{L}(\mathcal{H}_\otimes^g)$ then we can write (\ref{eq1}) as
\be
\sum\limits_i \Big( \hat{\mathbb{C}}^g_{ki} + \delta_{ik}\lambda^i(E) 
\Big) \xi_i^\alpha(E) = \Lambda^\alpha\xi_k^\alpha(E) \quad , 
\label{eq2}
\ee
which is a coupled system of eigenvalue equations in 
$\mathcal{H}_\otimes^g$ and its solutions $\xi_k^\alpha$ are the 
appropriate coefficients in the expansion (\ref{bo_ansatz}).\\
So far everything is exact and one could obtain the full set of 
solutions by first solving (\ref{schroedinger_phi}) and then use 
$\chi_i$ and $\lambda_i$ to compute the coefficients in the expansion 
(\ref{bo_ansatz}). However, even in our toy model we saw that there is 
in general no chance of obtaining analytic solutions. Quantum 
gravity is surely more difficult than a single molecule, so it would be 
surprising if one could make progress in the full theory without further 
simplifications. Thus, the crudest approximation is to 
use again the Born--Oppenheimer approximation 
which in this case amounts to neglecting all the off--diagonal 
terms in the matrix $\hat{\mathbb{C}}^g_{ki} $, that is we neglect the 
action of $\hat{P}$ on $\chi_i$. Then 
we can write $\braaaket{ \chi_k(\Phi;E)}{ \Big[ \hat{C}^g_{\rm 
kin}(\hat{P}, \hat{E}) + \hat{C}^g_{\rm pot}(\hat{E}) \Big] 
}{\chi_i(\Phi;E) }_\phi \approx \braaket{ \chi_k(\Phi;E)}{ 
\chi_i(\Phi;E) }_\phi \Big[ \hat{C}^g_{\rm kin}(\hat{P}, \hat{E}) + 
\hat{C}^g_{\rm pot}(\hat{E}) \Big] = \delta_{ki} \Big[ \hat{C}^g_{\rm 
kin}(\hat{P}, \hat{E}) + \hat{C}^g_{\rm pot}(\hat{E}) \Big]$. Thus we 
are left with
\be
\Big[ \hat{C}^g_{\rm kin}(\hat{P}, \hat{E}) + \hat{C}^g_{\rm 
pot}(\hat{E}) + \lambda^i(\hat{E}) \Big]\xi_i^\alpha(E)= 
\Lambda^\alpha\xi_k^\alpha(E) \quad . \label{bo_approx}
\ee
In the toy model we saw that using the pure Born--Oppenheimer 
approximation means that the influence of the light variables onto the 
quantum dynamics of the heavy variables is only effectively taken into 
account via their eigenvalues. This is also the case here: If we would 
set $\lambda^i(E) = 0$ then we would describe vacuum gravity and our 
solutions in the gravitational sector would not know anything about the 
matter content of the theory. One can regard (\ref{bo_approx}) as a 
first step towards quantum gravitational solutions which do not neglect 
the presence of matter fields, however matter is only `effectively' 
taken into account and not with respect to full dynamics. In a sense 
this is the quantum analog of the semiclassical Einstein equations 
$G_{\mu \nu} = \braket{ T_{\mu \nu} }_\phi$ where the expectation value 
of the stress energy tensor is used as a source in Einstein's equations. 
\vspace{.2cm}\\
At least in principle the strategy to solve dynamics in the {\it fully 
quantum gravitational setup} of the theory defined above would be the 
following:
\begin{enumerate}
\item \label{step1} As a first step  consider the Schr\"odinger equation 
(\ref{schroedinger_phi}) for the matter part. This can be interpreted as 
a fundamentally discrete quantum theory for matter degrees of freedom on 
a fixed background whose continuum limit would be quantum field theory 
(QFT) on a curved spacetime (CS). However, in the discrete setup we are 
not restricted to smooth metrics since the differential operators have 
been replaced by difference operators. Hence, $\hat{C}^\phi$ is a well 
defined 
operator for arbitrary values of $E_I^i(v)$ as long as it is non 
degenerate\footnote{In the degenerate case, QFT on CS methods cannot be 
used. By construction of our domain of definition, this case does not 
arise when restricting the action of the Hamiltonian operator to that 
domain.}. Thus, in a sense this can 
be seen as a generalisation of QFT from spacetimes with a curved 
smooth metric 
to ``spacetimes'' with a curved and even discontinuous metric. 
Nevertheless, if we 
consider $E_I^i(v)$ which approximate a smooth metric then 
(\ref{schroedinger_phi}) is a `lattice version' of QFT on CS. Note 
however, that the interpretation is quite opposite to the one usually 
employed for lattice theories: This theory of quantum gravity is 
fundamentally discrete and a continuum theory should be regarded as an 
approximation which is good for a certain set of states (the 
semiclassical ones) and bad for other states.\\
\item \label{step2} After analysing the matter dynamics on a fixed 
background one would obtain eigenfunctions $\chi_i(\Phi; E)$ and 
eigenvalues $\lambda_i(E)$ which are simply functions of $E$, i.e. it is 
not enough to consider QFT on CS for a single fixed background, but one 
would need knowledge about the whole class of quantum field theories on 
an arbitrary background.\\
Using these functions we could go ahead and analyse the gravitational 
sector of the theory, i.e. we would use equation (\ref{eq2}) to compute 
the coefficients $\xi_i^\alpha$ in ansatz (\ref{bo_ansatz}) and thus 
find solutions to the full quantum theory of gravity plus matter. 
However, from a practical point of view, there is no chance that one can 
exactly solve such a complicated system of coupled differential 
equations as (\ref{eq2}), so we must resort to approximation methods 
such as the pure 
Born--Oppenheimer approximation as described above. Thus, one would  use 
equation (\ref{bo_approx}) to calculate the coefficients 
$\xi_i^\alpha$. In this approximation the only imprint that matter 
leaves on the gravitational sector is the term proportional to 
$\lambda_i(E)$ in equation (\ref{bo_approx}). 

\end{enumerate}
However, from a practical point of view the situation is less clear: 
Even if we would succeed to somehow compute the eigenfunctions of the 
matter Hamiltonian $\chi_i(\Phi;E)$ (either numerically or using 
perturbative approaches such as an expansion in terms of Feynman 
diagrams) there still remains one big problem: In step \ref{step2} of 
the scheme described above we would need to solve the eigenvalue problem 
for the operator
\be
\hat{C}^g_{\rm kin}(\hat{P}, \hat{E}) + \hat{C}^g_{\rm pot}(\hat{E}) + 
\lambda^i(\hat{E}) \quad \label{born_simple}
\ee 
in order to get the correct coefficients $\xi_i^\alpha$. This operator 
is 
tremendously complicated, as we have seen in (\ref{C_g_q}) and even 
setting $\lambda_i(E)$ to zero does not substantially simplify the 
problem. For  $\lambda_i(E) = 0$ (\ref{born_simple}) is equivalent to 
the Hamiltonian constraint of a pure gravity theory on an algebraic 
lattice, and there is no hope that one can gain enough information about 
the space of its eigenfunctions in an analytic way. This is the reason 
why it is so complicated to incorporate the matter influence on the 
gravitational sector in the full theory on a more than formal level. 
Even considering $\lambda_i(E)$ just as a small perturbation does not 
help much: Without any knowledge about the spectrum of the unperturbed 
operator it is not meaningful to start a perturbative analysis.\\
At this point we want to stress that from a conceptual point of view 
there are no obstacles in the scheme we proposed above, it's just that 
the operator (\ref{born_simple}) is  {too} complicated without further 
simplifications. One could for example employ numerical methods to 
approximately solve (\ref{bo_approx}) in certain situations. A second 
option is to further exploit the fact that we want to describe 
situations where geometry can be treated almost classically, thus it 
makes sense to use a semiclassical approximation, which we will describe 
in the next paragraph.

\subsubsection{Born--Oppenheimer in the semiclassical regime}
\label{born}

What we are interested in is not the complete solution to the spectral 
problem. It would be sufficient to know the $\lambda_i(E)$ in the 
``neighbourhood'' of some three geometry $E_0$ and seek for coefficients
$\xi_i(E)$ that at the same time solve the eigenvalue equation, say 
in the pure Born -- Oppenheimer approximation, and in addition die off 
sufficiently quickly away from $E_0$. The problem is of course 1. to 
actually compute the $\lambda_i(E)$ in a suitable neighbourhood of some 
exactly solusolvableable $E_0$ with sufficient analytical control and 2. 
to find the eigenvectors $\xi_i$. 

As we are unable at present to say much about the solution of either of 
these problems we must resort here to a very crude approximation:\\
Pick any solvable $E_0$ (say Minkowski, FRW, ....) and choose as 
$\chi_i(E)$ a coherent state $\Psi_{Z_i}$ from the   
gravitational sector alone 
peaked on a classical phase space 
point $Z_J^j(v):= \frac{1}{\ell_P \sqrt{2}}\left( {E_J^j}_0- i {P^J_j}_i 
\right)$ (see section \ref{sec:classical_limit}). Notice that we pick
$\Re(Z_i)=E_0$ independent of $i$ but $\Im(Z_i)=P_i$ is not specified
at the moment. These states are 
kinematical coherent states, not dynamical ones, which means that they 
are not granted to be stable under the dynamics for a very long time.
However, when considering particle interactions the typical timescales 
on which such interactions happen are rather short, and at this stage we 
want to assume that the gravitational coherent states keep sharply 
peaked on a classical trajectory during these time intervals. This 
assumption seems natural from a particle physics point of view.\\
Clearly, the coherent states $\Psi_Z$ are not exact eigenstates of the 
Hamiltonian operator but at least approximately so. Moreover, they are 
orthogonal to a very good approximation whenever the $Z_i$ lie in 
different quantum cells of the phase space. As we will see, 
this condition will hold automatically whenever the $\lambda_i(E_0)$
differ sufficiently from each other. We now make the following Ansatz 
for the ``eigenstate'' of the physical Hamiltonian 
\be \label{2.1}
\Psi(E,\phi)=\sum_i\; c_i\; \Psi_{Z_i}(E)\;\chi_i(\phi;E_0)
\ee
for certain complex numbers $c_i$ (the $\Psi_{Z_i}$ themselves are 
normalised).
Notice that we have frozen the $E$ dependence of $\chi_i(\phi;E)$ at 
$E_0$ so that $C^g$ actually does not act on it. Therefore 
the Ansatz (\ref{2.1}) is not really of Born -- Oppenheimer type.
Inserting into the eigenvalue equation we obtain the exact equation
\be \label{2.2}
H\Psi=\sum_i 
\;c_i\;[(C^g+\lambda_i(E_0))\Psi_{Z_i}](E)\;\chi_i(\phi;E_0) 
=\Lambda\Psi
\ee
and thus due to the orthonormality of the $\chi_i$
\be \label{2.3}
c_i\;[(C^g+\lambda_i(E_0)-\Lambda)\Psi_{Z_i}](E)=0
\ee
for all $i$. Taking the inner product with $\Psi_{Z_j}$ we see that,
due to our assumption on the $\lambda_i(E_0)$ the equation for $j\not=i$ 
is trivially satisfied (approximately) while for $i=j,\;c_i\not=0$ we 
obtain 
\be \label{2.3}
<\Psi_{Z_i},C^g\Psi_{Z_i}>\approx 
C^g(E_0,P_i)=-\lambda_i(E_0)+\Lambda
\ee
which we use as a condition on the $P_i$ which therefore will mutually
different from each other as long as the $\lambda_i(E_0)$ are. Notice 
that our assumption $<\psi_{Z_i},\psi_{Z_j}>\approx \delta_{ij}$ is 
therefore self -- consistent.

However, as already stated, this procedure is unsatisfactory for 
several reasons of which we mention two: First it does not really follow 
the Born -- Oppenheimer
spirit and does not respect the true interaction between geometry and 
matter. This is also obvious from the fact that we nowhere used that the 
geometry operators were commuting, what we did could have been done 
in LQG as well and has already been done 
\cite{sahlmann_thiemann_curved1}.
The only new ingredient is that we took into account the 
backreaction in the sense of expectation values. Second, the above 
method can never reveal the spectral values of the physical Hamiltonian,
because we can always choose the $P_i$ to obey (\ref{2.3}) for any 
choice of $\Lambda$. 

Suffice it to say that a lot of work has to be invested in order to 
make the Born Oppenheimer Ansatz unfold its true power. To get more 
insight into the problem, we consider 
a truncation of the gravitational phase space (FRW spacetimes) in 
section \ref{sec:frw_example}.

\subsection{Construction of a Fock space}

Now we want to describe how one can in general construct a Fock space 
for a linear (i.e. the Hamiltonian is quadratic in the dynamical 
variables) classical theory. As explained above this situation emerges 
as step \ref{step1} in the Born--Oppenheimer analysis of our theory: 
When solving (\ref{schroedinger_phi}) and only considering matter 
dynamics the cotriads can be regarded as fixed non--dynamical functions 
because we have chosen a representation where they act as multiplication 
operators in the full theory. We follow largely 
\cite{sahlmann_thiemann_curved1} and then apply this 
scheme to the present situation. We assume the reader is familiar with the basic definitions 
concerning Fock spaces, for a more detailed review on the construction of 
Fock spaces in the context of QFT on CS we refer the reader to 
\cite{wald_qft}.
\vspace{.2cm}\\
Let $\HC_1$ be a Hilbert space (usually called the {\it one particle 
Hilbert space}) with inner product $\braaket{ \cdot}{ \cdot }_{1}$, 
$\psi = [\pi, \phi]$ a basis of $\HC_1$ and $\QC,\PC \in \LC(\HC_1)$ 
linear symmetric operators thereon (i.e. $\QC^\dagger = \QC, \PC^\dagger 
= \PC$). Assume further that their inverses $\QC^{-1}, \PC^{-1}$ exist 
on $\HC_1$. The Hamiltonian $H$ of a linear classical dynamical system 
can then be interpreted as a functional
\be
H: \HC_1 \rightarrow \RR; \; \psi = [\pi, \phi] \mapsto H(\psi) = 
\frac{1}{2}\braaaket{\pi}{ \PC }{\pi }_1 + \frac{1}{2} \braaaket{ \phi}{ 
\QC }{ \phi }_1 \label{matter_ham}
\ee
for given operators $\QC,\PC$.\\
Further assume that $\pi, \phi$ form a canonical pair, that is, the 
Hilbert space $\HC_1$ has also the structure of a symplectic 
manifold\footnote{We will omit the details here, but in the 
field--theoretic case one has to consider smeared versions of $\pi, 
\phi$ with appropriately chosen test functions to turn $\HC_1$ into an 
symplectic manifold.} (if $\dimm(\HC_1)$ is finite and $l, l'$ label its 
basis elements we have $\{ \pi_l, \phi_{l'} \} = \delta_{ll'}$ and 
$\{\pi_l, \pi_{l'} \} = \{\phi_l, \phi_{l'} \} = 0$ ). We are now aiming 
for a `quantisation' of that classical linear dynamical system, namely 
we want to construct a Hilbert space $\HC$ which carries a 
representation of the classical observable algebra given by the Poisson 
algebra of $\pi$ and $\phi$.\\
One convenient way of doing this, which naturally allows for an 
interpretation in terms of `particles' is the construction of a Fock 
space: That is, we are implementing a unitary map $U$ on  $\HC_1$  and 
use this map to construct complex functions $z(\pi, \phi)$ such that $\{ 
\bar{z}_l, z_{l'}  \} = i \delta_{ll'}$ where $\bar{z}$ is the complex 
conjugate to $z$. But this is just the commutator algebra of creation 
and annihilation operators and it is well known that such an algebra can 
naturally be represented on $\FC_s(\HC_1)$, the symmetric Fock space 
over $\HC_1$.\\
Let us start with the Hilbert space $\HC_1$ and complexify it. Now 
choose a subspace $\HC_1^{\CC +} \subset \HC_1^{\CC}$ such that there 
exists a unitary, real--linear map $U: \HC \mapsto \HC_1^{\CC+}$. Using 
this map we can rewrite the Hamiltonian in the  form
\be
H = \braaaket{ \bar{z}}{ \omega }{z }_{\HC_1^{\CC+}} \label{matter_hamz}
\ee
where $z \in \HC_1^{\CC+}$ and $\bar{z}$ is the complex conjugate to $z$ 
in $\HC_1^{\CC+}$ and $\omega \in \LC(\HC_1^{\CC+})$ as defined below. 
$z$ is obtained by using this unitary map $U$ through\footnote{It will 
become obvious why the operator $\DC$ must have this rather complicated 
form in the next paragraph.}
\be
z & := & \frac{1}{\sqrt{2}}U (\DC \phi - i 
\DC^{-1}\pi)\label{annihilation_op}\\
\DC & := & [\PC^{-1/2}(\PC^{1/2}\QC\PC^{1/2})^{1/2}\PC^{-1/2}]^{1/2} 
\quad ,
\ee
and $\omega$ is given through
\be
\omega := U \DC^{-1}\QC\DC^{-1}U^{-1} \quad
\ee
Note that $\DC$ is a symmetric operator with well defined inverse if 
$\QC$ and $\PC$ are.\\
Then one gets
\be
\braaaket{\bar{z}}{ \omega }{ z }_{\HC_1^{\CC+}} & = & 
\frac{1}{2}\braaaket{ U(\DC\phi + i\DC^{-1}\pi)}{ 
U\DC^{-1}\QC\DC^{-1}U^{-1}U}{ (\DC\phi - i \DC^{-1}\pi)}_{\HC_1^{\CC+}} 
\nn \\
& = & \frac{1}{2} \braaket{ \DC\phi + i\DC^{-1}\pi}{ \DC^{-1}\QC 
\DC^{-1}(\DC\phi - i \DC^{-1}\pi)}_1 \nn \\
& = & \frac{1}{2}\braaket{ (\DC^{-1})^\dagger \DC \phi}{ \QC \phi}_1 + 
\frac{1}{2}\braaket{  \pi}{ (\DC^{-1})^\dagger \DC^{-1}\QC 
\DC^{-1}\DC^{-1} \pi }_1\nn \\
& & + \frac{i}{2}\braaket{ \pi}{ (\DC^{-1})^\dagger \DC^{-1} \QC \phi  
}_1 - \frac{i}{2}\braaket{ (\DC^{-1})^\dagger (\DC^{-1})^\dagger 
\QC^\dagger (\DC^{-1})^\dagger \DC \phi}{ \pi }_1 \nn \\
& = & \frac{1}{2}\braaket{ \phi}{ \QC\phi}_1 + \frac{1}{2}\braaket{ 
\pi}{ \DC^{-2}\QC \DC^{-2}\pi}_1 \nn \\
& & + \frac{i}{2}\braaket{ \pi}{ \DC^{-2}\QC \phi}_1 - 
\frac{i}{2}\overline{\braaket{ \pi}{ \DC^{-2}\QC \phi }_1} \nn \\
& = & \frac{1}{2}\braaket{ \phi}{ \QC \phi}_1 + \frac{1}{2}\braaket{  
\pi}{ \PC \pi}_1 \nn \\
& = & H
\ee
where in the second equality we used that $U$ is a unitary map, in the 
fourth equality that $\QC,\PC$ and $\DC$ are symmetric and in the fifth 
equality that $\DC$ is real and $\DC^{-2}\QC\DC^{-2} = \PC$. This last 
equality can be seen by explicitly writing $\DC$ in terms of $\QC,\PC$ 
and then multiplying both sides first by $\PC^{-1/2}$ and then by 
$(\PC^{1/2}\QC\PC^{1/2})^{1/2}$.\\
From equation (\ref{annihilation_op}) one can verify that $\bar{z}, z$ 
indeed form the Poisson algebra of latter operators and therefore the 
Hamiltonian (\ref{matter_ham}) can be defined as a linear operator on 
$\FC_s(\HC_1^{\CC+})$ in the form (\ref{matter_hamz}).\\
\\
In our case the one particle Hilbert space is given by $\HC_1 = 
l_2(V(\alpha))$, the space of square--summable functions over the set of 
vertices $V(\alpha)$. The inner product on this space is simply given by 
$\langle \cdot , \cdot \rangle_1: \HC_1 \times \HC_1 \rightarrow 
\CC;\quad f,f' \mapsto \sum\limits_{v\in V(\alpha)} \bar{f}(v)f'(v)$. 
Field coordinates are given by $\pi(v), \phi(v)$ and these form a 
canonical pair with the Poisson bracket $\{ \pi(v), \phi(v') \} = 
\delta_{vv'}$. If we consider the cotriads on each vertex as prescribed 
functions then the ``classical'' Hamiltonian is given by
\be
H & = & \frac{1}{2}\sum\limits_{v \in V(\alpha)} \frac{1}{\det 
(E)(v)}\pi^2(v) + \det (E)(v) \phi(v) (-\Delta + m^2) \phi(v)     \quad 
,
\ee
with the discrete Laplace--Beltrami operator $\Delta$ given by
\be
\Delta = \frac{1}{2 \det (E)(v)} \nabla^+_I [\det (E)(v) \epsilon^{IKL} 
\epsilon^{JMN} E_K^k(v) E_L^l(v) E_M^k(v) E_N^l(v) \nabla^-_J ]
\ee
with the forward and backward derivative, $\nabla^+_I$ and $\nabla^-_I$, 
acting on functions $f \in l_2(V(\alpha))$ as $\nabla^+_I f(v) := f(v+I) 
- f(v)$ and $\nabla^-_I f(v):= f(v)-f(v-I)$ respectively.
Thus $\QC$ and $\PC$ are given as linear symmetric operators on $\HC_1$ 
as
\be
\QC & = & \frac{1}{\det (E)(v)} \quad ,\\
\PC & = & \det (E)(v) (-\Delta + m^2) \quad .
\ee
So we can construct $\bar{z},z$ such that they have the desired Poisson 
brackets and we can represent them as creation and annihilation 
operators on the Fock space $\FC_s(l_2(V(\alpha)))$.\\

\subsection{A minisuperspace example: QFT on 
FRW--spacetimes}\label{sec:frw_example}

We want to illustrate the Born Oppenheimer Decomposition as well as its 
semiclassical approximation  
explained in section \ref{born} 
for a case for which we have sufficient mathematical control over the 
equations. We will choose a ``hybrid model'', that is,  homogeneous, 
isotropic FRW -- 
minisuperspace model coupled to inhomogeneous matter which recently 
was advocated in the context of Gowdy models \cite{Mercedes}. 
That is, in the 
language of the previous section,
we pick as a spatial geometry a homogeneous cotriad $E_0$ and pick as a 
neighbourhood around it all homogeneous cotriads $E$. This neighbourhood 
confines us to the homogeneous sector of GR while matter is treated as a 
QFT. In the classical theory this leads to consistent equations of 
motion because the equations of motion for geometry only depend on the 
integrated matter fields.

We will assume that the spatial topology of the 
universe which arises in the classical limit of this quantum theory is 
given by $\Sigma = T^3$. In the classical continuum theory for these 
model the cotriad is given by
\be
E_a^i (\sigma) :=  E \delta_a^i
\ee
where $E$ is a homogeneous function of time only, related to the  scale 
factor by $a=|E|$. Its canonical momentum
\be
P_i^a (\sigma) := P \delta_i^a
\ee
is also homogeneous and, in terms of $\tau$--derivatives takes the form 
$P = -4E\dot{E}$. Assuming that these variables emerge from some 
averaging procedure
\be
P := \frac{1}{3V_0}\int\limits_{T^3}d^3\sigma P^a_i(\sigma)\delta_a^i 
\nn \\
E := \frac{1}{3V_0}\int\limits_{T^3}d^3\sigma E_a^i(\sigma)\delta^a_i
\ee
with $V_0$ the coordinate volume of $T^3$ they have canonical Poisson 
brackets
\be
\{ P , E \} = \frac{\kappa}{3 V_0}
\ee
In the FRW -- minisuperspace model the  physical Hamiltonian $H_{\rm 
phys}$ reduces to 
$H_{\rm phys}=\frac{1}{3V_0}\int_{\cal 
T^3}d\sigma\sqrt{(C^g+C^\phi)^2(\sigma)}=C^g+C^\phi$ because the spatial 
diffeomorphism constraint identically vanishes.
Inserting the ansatz for the elementary phase space variables in the 
general expression (\ref{C^g}) the gravitational part of the integrated 
Hamiltonian is simply
\be
C^g = -\frac{3V_0}{8\kappa}\frac{P^2}{|E|}
\ee
Later we will consider spacetimes with non--vanishing cosmological 
constant $\Lambda \neq 0$, in this case the gravitational part of the 
Hamiltonian constraint is given by
\be
C^g_\Lambda = -\frac{3V_0}{8\kappa}\frac{P^2}{|E|} + 
\frac{1}{\kappa}\Lambda V_0 |E|^3 \quad .
\ee
Passing to the quantum theory this means we have operators 
$\hat{E}_I^i(v) = \hat{E}\delta_I^i$ and $\hat{P}^I_i(v) = 
\hat{P}\delta_i^i$ which do not depend on the algebraic lattice point 
$v$, $\hat{E} = E \cdot$ acts as a multiplication operator and $\hat{P} 
= i\ell_P^2 \partial_E$ as a derivative operator. As in the Wheeler--de 
Witt theory the Hilbert space of this minisuperspace model is given by 
$\HC = L_2(\RR)$, the space of square--integrable functions over the 
real line. In the absence of matter fields dynamics in the gravitational 
sector is then generated by an operator $\hat{C}^g_\Lambda$ that is 
obtained from the classical expression by replacing $E \rightarrow 
\hat{E}, P \rightarrow \hat{P}$ in some suitable operator ordering. The 
solutions $e_\lambda(E)$ to the eigenvalue problem
\be
\hat{C}^g_\Lambda e_\lambda(E) = \lambda e_\lambda(E)
\ee
can be given in terms of some hypergeometric functions. If we choose a 
symmetric operator ordering the eigenvalue problem reads
\be
-\frac{3 V_0 \ell_P^4}{8\kappa}E^2 \frac{\partial^2}{\partial E^2} 
e_\lambda(E) + \frac{3 \kappa V_0 \ell_P^4}{8} |E| \frac{\partial 
}{\partial E}e_\lambda(E) - \frac{V_0 \Lambda}{\kappa}E^6 e_\lambda(E) + 
\lambda |E|^3 e_\lambda(E) = 0
\ee
and the solutions $e_\lambda(E)$ can explicitly be written in terms of 
Kummer's functions of the first and second kind (see 
\cite{abramowitz_stegun} for their properties). For the special case 
$\Lambda = 0$, $\lambda = 0$ the solutions reduce to ordinary Bessel 
functions and in \cite{kiefer_wave_packets} it was shown that in this 
basis one can construct `wave packet solutions' whose quantum evolution 
stays sharply peaked 
on a classical trajectory in phase space. Thus, these quantum solutions 
are close to classical ones not only at one instant of time but also 
under dynamical evolution. We will follow here a somewhat different 
route.

On such a FRW--spacetime one can  explicitly carry out the construction 
of a Fock space using the method explained in the last section: If we 
want this theory to describe quantum field theory on an FRW universe 
with spatial topology $\Sigma = T^3$, the number of vertices of $\alpha$ 
should be finite. Assuming further that the number of vertices in each 
spatial direction is $N$ we have $|V(\alpha)| = N^3$. 
Let $\Phi(v)$ and $\Pi(v)$ be the scalar field and its canonical 
momentum at each vertex $v$ of the algebraic graph. Then the classical 
Hamiltonian for a scalar field on a space with cotriad $E$ is given by
\be
C^\phi := \frac{1}{2}\sum\limits_{v \in V(\alpha) } 
\Pi(v)\frac{1}{|E|^3}\Pi(v) + \Phi(v) |E|^3\left[ \frac{-\Delta_{\rm 
flat}}{E^2} + m^2 \right]\Phi(v) \label{starting_qft}
\ee
with the ordinary flat lattice Laplacian $\Delta_{\rm flat } := 
\delta^{IJ}\nabla^-_I \nabla^+_J$.
Using spatial homogeneity and flatness one can introduce a discrete 
Fourier transform from $V(\alpha)$ to $K(\alpha)$, the associated 
momentum space. $K(\alpha)$ is discrete (because $V(\alpha)$ is bounded) 
and bounded (because $V(\alpha)$ is discrete). If we choose each of the 
dimensions in $T^3$ to be of unit coordinate length and choose to embed 
$V(\alpha)$ uniformly in $T^3$, then the vertices are labeled by triples 
$v = (v_1, v_2, v_3)$ with $v_1, v_2, v_3 \in \{\frac{1}{N}, \frac{2}{ 
N}, \dots,  \frac{N}{N} \}$. Thus, for each function $f(v)$ we can 
define its Fourier transform
\be
f_k := \sum\limits_{v} f(v)e^{-ik\cdot v}
\ee
with $k = (k_1, k_2, k_3)$ and $ k_1, k_2, k_3 \in \{ 2 \pi, 4 \pi, 
\dots, 2 N \pi \} =: K(\alpha)$. The inverse transformation is then 
given by
\be
f(v) = \frac{1}{N^3}\sum\limits_{k \in K(\alpha)} f_k e^{+ik\cdot v} 
\quad .
\ee
Furthermore we have the identities $\frac{1}{N^3}\sum\limits_{k}e^{k 
\cdot (v - v')} = \delta_{v,v'}$ and $\frac{1}{N^3}\sum\limits_{v}e^{(k 
- k') \cdot v } = \delta_{k,k'}$ where the deltas on the right hand side 
are both Kronecker deltas.\\
One can show that the Fourier transform of the flat Laplacian acting on 
a function $f(v)$ is given by
\be
\Delta_{\rm flat} f(v) := \frac{1}{N^3}\sum\limits_{k \in 
K(\alpha)}\left[-4 N^2 \sum\limits_{I=1,2,3} \sin^2(\frac{k_I}{2 N})   
\right] f_k e^{+i k \cdot v} \quad ,
\ee
which in the continuum limit just gives the usual expression 
$\mathop{lim}\limits_{N\rightarrow \infty}\Delta_{\rm flat} f_k = -k^2 
f_k$. Here we chose a naive discretisation for implementing the 
derivative on the lattice. Improved discretisations can be found in the 
context of perfect actions \cite{Hasenfratz} that provide a framework in 
which discretisation artefacts can be avoided. For this article we will 
restrict our discussion to the standard (naive) discretisation procedure 
and will present the discussion on improved actions elsewhere.\\
Thus, we can use the Fourier transform and write the Hamiltonian as
\be
C^\phi = \frac{1}{2 N^3}\sum\limits_{k \in K(\alpha)}\Pi_k 
\frac{1}{|E|^3}\Pi_{-k} + \Phi_k |E|^3 \left[ 
\frac{4N^2}{E^2}\sum\limits_{I=1,2,3}\sin^2(\frac{k_I}{2 N}) + m^2 
\right]\Phi_{-k}.
\ee
In Fourier space the symplectic structure is given by $\{\Pi_k, 
\Phi_{k'}  \}  = N^3 \delta_{k, -k'}$ where $-k \stackrel{!}{=} (2 \pi N 
- k)$.
Introducing the Hilbert space $\HC_1 := l_2(K(\alpha))$ with inner 
product $\braaket{f}{g}_k := \frac{1}{N^3}\sum\limits_{k \in K(\alpha)} 
f_k g_{-k}$ we can write
\be
C^\phi = \frac{1}{2}\braaaket{\Pi}{\PC}{ \Pi}_k + 
\frac{1}{2}\braaaket{\Phi}{\QC}{ \Phi}_k
\ee
with $\PC := \frac{1}{E^3}$ and $\QC:= |E|^3 \left[ 
\frac{\gamma^2(k)}{E_0^2} + m^2 \right]$ with $\gamma^2(k) := 4N^2 
\sum\limits_{I=1,2,3}\sin^2(\frac{k_I}{2 N})$ both simply acting by 
multiplication.\\
Here we are in the special situation that $[\PC,\QC]=0$ as linear 
operators on $\HC_1$, thus we can explicitly construct the algebra of 
creation and annihilation variables: For commuting $\PC$ and $\QC$ we 
get
\be
\DC  = \left[ \frac{\QC}{\PC} \right]^{1/4} = |E|^{3/2} \left[ 
\frac{\gamma^2(k)}{E^2} +  m^2 \right]^{1/4}
\ee
and thus
\be
z_k := \frac{1}{\sqrt{2}}\left[ |E_0|^{3/2} \left[ 
\frac{\gamma^2(k)}{E^2} +  m^2 \right]^{1/4}   \Phi_k - i |E|^{-3/2} 
\left[ \frac{\gamma^2(k)}{E^2} +  m^2 \right]^{-1/4}\Pi_k \right]
\ee
One can easily check that $\{ \bar{z}_k, z_{k'} \} = i \delta_{k,-k'}$. 
The standard representation of this algebra is as creation and 
annihilation operators on the Fock space $\FC(\HC_1)$.\\
Using these variables the Hamiltonian is given by
\be
C^\phi = \braaaket{\bar{z}}{\omega_k}{z}_k
\ee
with $\omega_k = \sqrt{ \frac{\gamma^2(k)}{E^2} +  m^2 }$. Compared to 
standard QFT on Minkowski space, where the spectrum takes values 
$\omega_k = \sqrt{k^2 + m^2}$ there are two differences: First, we are 
considering a discrete theory, thus the momentum squared gets replaced 
by $\gamma^2(k)$, a bounded expression that converges against $k^2$ in 
the limit $N \rightarrow \infty$. Second, the 
$z_k,\;\omega_k$ and therefore the vacuum $|0>$ depends explicitly on 
$E$.\\
If we choose this representation and denote by $\hat{a}_k^\dagger, 
\hat{a}_k$ the corresponding ladder operators in the Fock space can be 
characterised through a cyclic vector (`the vacuum') $\ket{0}$ defined 
through the condition $\hat{a}_k \ket{0} = 0 \quad \forall k$. All other 
elements of $\FC$ can then be constructed by letting the creation 
operators $\hat{a}^\dagger$ act on $\ket{0}$ and the Hamiltonian 
operator is given by
\be
\hat{C}^\phi := \sum_k\; \omega_k\; \hat{a}_k^\dagger\; 
\hat{a}_k
\ee
For the vacuum state $\ket{0}$ the expectation value of the Hamiltonian 
is simply $\braaaket{0}{\hat{H}}{ 0} = 0$\footnote{Here the inner 
product $\braaket{.}{.}$ is taken on the full Fock space $\FC$ as 
opposed to the one--particle inner product $\braaket{.}{.}_k$} and if we 
denote $n$--particle states for particles with momenta $k_1, \dots, k_n$ 
by
\be
\ket{n} := \hat{a}^\dagger_{k_1} \dots \hat{a}^\dagger_{k_n}\ket{0}
\ee
we get
\be
\hat{H}\ket{n} = 
[\sum\limits_{i=1}^{n}\omega_{k_i}]\;\ket{n}=:\lambda_n(E)\;|n>
\ee
where the role of the index $n$ is played by the labels $k_1,..,k_n$.
Let us now analyse the coupled quantum system gravity (in the 
minisuperspace approximation) plus matter according to the scheme we 
proposed in section \ref{born}: To ensure that we have nontrivial 
dynamics for the gravity part already in the absence of matter we choose 
to work with the Hamiltonian $C^g_\Lambda$ with non--vanishing 
cosmological constant.\\
\\
In order to carry out the Born Oppenheimer programme, we have to 
construct states $\Psi(E, \Phi)= 
\sum\limits_n \xi(E) \ket{n}(E, \Phi) $ and to compute 
the coefficients $\xi(a)$ in the Born--Oppenheimer ansatz. 
It is convenient to first pass to a new set of canonical variables which 
are adapted to the specific form $C^g_\Lambda$ and then to quantise the 
system in these new variables for which coherent states can be more 
easily constructed. We introduce the following variables
 \be
 \label{Patilde}
 \tilde{P}_:=\sqrt{3V_0}\frac{P}{\sqrt{|E|}},\quad 
\tilde{E}:=\sqrt{3V_0}\frac{2}{3}\sgn(E)|E|^{\frac{3}{2}},\quad{\rm 
then}\quad \{\tilde{P},\tilde{E}\}=\kappa
  \ee
 and all other Poisson brackets are vanishing. Here we neglected terms 
involving a derivative of the signum function since these terms only 
contributes when $E=0$. The quantity $\tilde{P}$ can be defined as long 
as $E\not=0$ which is given in the classical theory. In the quantum 
theory we will modify the corresponding Schwartz functions as before by 
means of the function $s_\epsilon(\tilde{E})$. In these variables 
$C^g_\Lambda$ takes the form
 \be \label{sign}
 C^g_\Lambda = -\frac{1}{8\kappa}\tilde{P}^2 + \frac{3}{4\kappa}\Lambda 
\tilde{E}^2
 \ee
which is almost of form of a harmonic oscillator except that it has the 
wrong sign in the kinetic term.
We now choose a representation on $L_2(\RR, d\tilde{E})$ in which $P$ 
is implemented as $-i\ell_P^2\partial/ \partial\tilde{E}$ and 
$\tilde{E}$ is represented as a multiplication operator.
  The quantum Hamiltonian is thus
  \be
 \hat{C}^g_\Lambda = 
\frac{\ell_P^4}{8\kappa}\frac{\partial^2}{\partial\tilde{E}^2} + 
\frac{3}{4\kappa}\Lambda \hat{\tilde{E}}^2 
 \ee
This motivates to construct coherent states which simply those of that
harmonic oscillator which results when switching to a positive sign in 
front of 
the kinetic term in \ref{sign}. It is then obvious that 
these coherent states approximate the operators $\tilde{P}$ 
and $\tilde{E}$ semiclassically well i.e.
\be
\langle \tilde{\Psi}^\epsilon_{Z_0}\, ,\, \tilde{E} 
\tilde{\Psi}^\epsilon_{Z_0}\rangle 
=\tilde{E}_0=\sqrt{3V_0}\frac{2}{3}\sgn(E_0)E_0^{\frac{3}{2}}, \quad
\langle \tilde{\Psi}^\epsilon_{Z_0}\, ,\, \tilde{P} 
\tilde{\Psi}^\epsilon_{Z_0}\rangle 
=\tilde{P}_0=\sqrt{3V_0}\frac{P_0}{\sqrt{|E_0|}}
\ee
as long as we are not close to a classical singularity $E_0=0$ where  
the 
function $s_\epsilon$ is constant, see also discussion in section 
\ref{sec:classical_limit}.
Therefore these states will also give the correct classical value for 
the expectation value of $C^g_\Lambda$. \\
\\
The exact Born Oppenheimer programme now consists in the following:
We are looking for the (generalised) eigenvalues $\mu$ and the system of
corresponding eigenfunctions $n\mapsto \xi_n(E)$ of the equation 
\be 
[C^g_\Lambda+\lambda_n]\xi_n=\mu \chi_n
\ee
where $\lambda_n$ is a multiplication operator. This can now be done 
by using quantum mechanical perturbation theory: Multiply this equation 
by $\kappa$ and treat $\kappa C^g_\Lambda$ as the ``free'' Hamiltonian 
and $\kappa \lambda_n$ as a perturbation potential (notice that $\kappa$ 
is an extremely small parameter. Assuming that the spectral problem for 
$C^g_\Lambda$ has been solved we can then compute the $\mu$ and the 
corresponding systems $\xi_n$. This is beyond the scope of the present 
paper and we plan to come back to this problem in a future 
publication. From the resulting eigenfunctions $\Psi_\mu(E,\phi)$ 
one then has to select those which are peaked on a given spatial 
geometry $E_0$ and a given extrinsic curvature $P_0$. \\
The approximate (semiclassical) programme consists in choosing coherent 
states
$\chi_n:=\psi_{Z_n}$ with $\Re(Z_n)=E_0,\;\Im(Z_n)=P_n$ for some given 
$E_0$ and to 
satisfy the eigenvalue equation in the sense of expectation values only 
which results in conditions which are approximately given by 
\be 
C^g_\Lambda(E_0,P_n)+\lambda_n(E_0)=\mu
\ee
which in this case can be trivially solved for $P_n$.

\section{Conclusions and Outlook} \label{sec:conclusions}

In this article we investigated the question of how one can understand 
(matter) QFT on a fixed background starting from first principles, i.e. 
from a theory where gravitational and matter degrees of freedom are 
treated as quantum variables. In contrast to ordinary QFT on (curved) 
spacetimes such a theory describes quantum matter on a quantum 
spacetime.
We performed an analysis using Born--Oppenheimer methods and it turns 
out that (matter) QFT on a fixed background emerges as an intermediate 
step when trying to find solutions for the full quantum theory.
Within the Born--Oppenheimer approximation one assumes that the change 
in spacetime geometry is very small on typical timescales of particle 
interactions. This is obviously the case in situations where one expects 
ordinary QFT to be valid but will be violated in the Planck scale regime 
where the notion of a smooth background spacetime is not appropriate any 
longer because quantum fluctuations of the quantum geometry can not be 
neglected even for very short timescales. 
However, within its regime of applicability, this assumption means that 
the  geometrical operators, which encode the gravitational degrees of 
freedom, can be well approximated by their classical counterparts as 
long as we are only interested in matter dynamics.\\
 Classically, gravity couples via the spatial metric to standard matter 
(at least for scalar fields and gauge bosons) and not via its canonical 
momentum.
 In order to be able to apply the Born--Oppenheimer approximation scheme 
we need to require that the operator analog of the spatial metric can be 
well approximated by the classical metric in some regime. This is surely 
the case when geometrical quantities are represented as multiplication 
operators, but is not obvious when these  quantities are represented as 
derivative operators because the derivative operators might not commute 
among each other. \\
In ordinary LQG we are exactly in the latter situation: The flux 
operators, which are the quantum versions of densitised triads smeared 
along 2--dimensional surfaces, are represented as derivative operators 
on the Ashtekar--Lewandowski Hilbert space $\HC_{\rm AL}$ and 
furthermore form a non--commuting operator algebra. Thus it is not 
immediately obvious in which sense they can be used to approximate a 
classical geometry in the sense of the Born--Oppenheimer 
approximation.

This led us to consider a new algebra of operators and a representation 
thereof  which 
is similar to ordinary LQG in the sense that its Hilbert space carries a 
basis of states which are defined on certain graphs but deviates from 
LQG in the choice of representation: We worked in a representation where 
cotriads are represented as ordinary multiplication operators and their 
canonical momenta as derivation operators.  By this we obtain a 
theory of quantum gravity defined on an algebraic graph to which we can 
directly 
apply the  
Born -- 
Oppenheimer methods  because the matter degrees of freedom now couple to 
ordinary multiplication operators. Since the quantum theory defined in 
this article is an AQG -- inspired quantisation on a fixed algebraic 
graph where the notion of a manifold and therefore a classical spacetime 
has a priori no interpretation, we need to analyse its semiclassical 
sector in order to be able to make contact to ordinary QFT.
\\
Using these methods we saw how QFT on a fixed background emerges out of 
a full quantum theory and moreover were able to give a conceptually 
framework that in principle allows to calculate backreactions from 
quantum 
matter 
onto the quantum dynamics within the gravitational sector.
\\
However, from a practical point of view there is little chance that one 
will be able to perform these calculations analytically as they require 
a substantial amount of input concerning the quantum dynamics from the 
gravitational sector. This means that in order to be able to fully 
understand the problem and to be able to calculate corrections to 
particle physics processes from first principles one will need to get 
more control over the solution space of the gravitational Hamiltonian.\\
As in previous contributions to this subject, there exists a
`semiclassical route', 
i.e. one can consider states that can be written as a tensor product of 
certain appropriately defined Fock states for the matter sector on a 
fixed background and 
coherent states (which are sharply peaked on a classical geometry 
around this background) for 
the gravitational sectors. However, while backreaction effects can 
be taken into account in the sense of expectation values, these states 
are not eigenstates of the quantum Hamiltonian, therefore do not shed 
light on the true spectral problem and in particular do not really 
follow the Born -- Oppenheimer scheme.\\
Our analysis aims at a complete 
understanding of particle scattering processes within a full theory of 
quantum gravity for which some knowledge on the spectrum of the physical 
Hamiltonian and in particular its vacuum sector might prove useful. 
In order to illustrate our method we tested the framework using a FRW 
minisuperspace model for this quantisation in section 
\ref{sec:frw_example}. There we could see more explicitly how the 
presence of quantum 
matter fields can be taken into account into the dynamics of the 
gravitational sector using the Born--Oppenheimer method.\\
Of course, in principle there is no obstruction to use this method to 
analyse how a 
quantum field theory for matter emerges out of a quantum gravitational 
setting as loop quantum cosmology. Contrary to full LQG the geometrical 
operators are represented as multiplication operators in loop 
quantum cosmology as well and it will be interesting to study the 
relation\footnote{LQC is a based on a representation that mimics 
that of LQC, it is not of Schr\"odinger type in contrast to the 
representation used in this paper.} 
between our method and the results obtained in 
\cite{AshtekarKaminskiLew}.\\
It is challenging to generalise the Born Oppenheimer method to 
representations for which the three geometry is 
represented by non commuting operators such as the non--commuting flux 
operators of LQG. But this requires a completely new input.
}
$\;$\\
\vspace{1cm}
\\
{\large\bf Acknowledgements}\\
\\
JT wants to thank Aristide Baratin, Carla Cederbaum, Bianca Dittrich and 
Hanno Sahlmann for valuable discussions.
KG wants to thank Stefan Hofmann for discussions.
The part of the research performed at the Perimeter Institute for
Theoretical Physics was supported in part by funds from the Government 
of
Canada through NSERC and from the Province of Ontario through MEDT.

\appendix

\section{Infinite tensor product Hilbert spaces} \label{ITP}

Infinite tensor product (ITP) Hilbert spaces emerge as a generalisation 
of the tensor product of a finite number of Hilbert spaces. They were 
already studied by von Neumann in the thirties and we want to give a 
brief summary of the material that is relevant for our purposes. For a 
more detailed discussion see for instance \cite{GCSIV} or von Neumann's 
original article \cite{neumann_itp}.\\
Let $\HC_i$ be Hilbert spaces and $i \in \IC$ for some index set $\IC$. 
In general one can allow for arbitrary cardinality of $\IC$ but for our 
purposes it will be enough to consider $|\IC| = \aleph$.\\
The ITP Hilbert space $\HC_\otimes$ is defined as the closure of the 
finite linear span of vectors of the form $\otimes_f := \otimes_i f_i$ 
for $f_i \in \HC_i$ with respect to the inner product
\be
\langle \otimes_f, \otimes_f' \rangle_{\HC_\otimes} := \prod\limits_{i} 
\langle f_i, f'_i \rangle_{\HC_i} \quad.
\ee
Here the infinite product $\prod\limits_i z_i$ of complex numbers $z_i = 
|z_i| \exp (i \phi_i)$ is defined via
\be
\prod\limits_{i}z_i := \left[ \prod\limits_i |z_i|  \right] \exp\left( i 
\sum\limits_i \phi_i \right)
\ee
if the absolute value $\prod\limits_i |z_i|$ and the phase 
$\sum\limits_i \phi_i$ both converge. In this case we call $z := 
\prod\limits_{i}z_i$ convergent. Otherwise we set $\prod\limits_i 
z_i=0$. If the absolute value $\prod\limits_{i \in \IC} |z_i|$ converges 
but not necessarily the phase, we say that $\prod\limits_{i \in \IC}z_i$ 
is {\it quasi convergent}.
One can show that for $z=\prod\limits_i z_i \neq 0$ given any $\delta > 
0$ there exists a finite subset $\IC_\delta \subset \IC$ such that $|z - 
\prod\limits_{i \in \IC_\delta}| < \delta$. We will only be interested 
in elements $\otimes_f$ that have non--vanishing norm.\\
On $\HC_\otimes$ one can define different notions of equivalence: Two 
vectors $\otimes_f$ and $\otimes_{f'}$ are said to be {\it strongly 
equivalent} if and only if $|\sum\limits_{i \in \IC}\langle  f_i, f'_i   
\rangle_{\HC_i} - 1|$ converges. Strong equivalence classes of vectors 
$\otimes_f$ will be denoted by $[f]$. It follows that $\langle 
\otimes_f, \otimes_{f'} \rangle_{\HC_\otimes} = 0$ if either $[f] \neq 
[f']$ or $[f] = [f']$ and $\langle f_i, f'_i \rangle_{\HC_i} = 0$ for at 
least one $i$.\\
If we set $(z \cdot f)_i := z_i f_i$ then $\otimes_{z \cdot f} = 
(\prod\limits_{i \in \IC}) \otimes_f$ fails to hold if $\prod\limits_{i 
\in \IC} z_i$ is not convergent. Provided that there exists such $z$ we 
say that $\otimes_f$ and $\otimes_{f'}$ are {\it weakly equivalent} if 
and only if $[z\cdot f] = [f]$. These weak equivalence classes will be 
denoted by $(f)$. One can show that to functions $\otimes_f, 
\otimes_{f'}$ are in the same weak equivalence class if and only if 
$\sum\limits_{i \in \IC}| \; \ |\langle f_i, f'_i \rangle_{\HC_i} | \; - 
1| $ converges. Hence, strong equivalence implies weak equivalence.\\
One can show that the closure of the span of all vectors in the same 
strong equivalence class $[f]$, denoted by $\HC^\otimes_{[f]}$ is 
separable, consisting of the completion of the finite linear span of 
vectors of the form $\otimes_{f'}$ where $f'_i = f_i$ for all but 
finitely many $i$. The whole ITP Hilbert space is the direct sum (in 
general an infinite sum) of these separable subspaces, $\HC_\otimes = 
\sum\limits_{[f]}\HC^\otimes_{[f]}$. Let also $\HC^\otimes_{(f)}$ be the 
closure of the finite linear span of vectors $\otimes_{f'}$ with $(f') = 
(f)$. Then the strong equivalence subspaces of $\HC^\otimes_{(f)}$ are 
unitarily equivalent with corresponding unitary operators of the form 
$U_z\otimes_f := \otimes_{z \cdot f}$ with $\prod\limits_{i \in \IC} 
z_i$ quasi convergent.\\
We see that although the ITP Hilbert space $\HC_\otimes$ is not 
separable and might look ``too large'' at first sight, it decomposes 
into a direct sum of separable Hilbert spaces $\HC^\otimes_{[f]}$ when 
taking into account strong equivalence classes of vectors $[f]$. In QFT 
on a fixed curved background the theory is naturally formulated on a 
separable Hilbert space, one for each background spacetime. So in a 
sense these ITP Hilbert spaces emerge as a natural generalisation when 
going from QFT on CS to quantum gravity, which should incorporate {\it 
QFT on all possible backgrounds simultaneously}.

\bibliographystyle{utphys}
\bibliography{thebibliography}

\end{document}